\documentclass{article}
\usepackage{url} 
\usepackage{lineno}

\usepackage{array}
\usepackage{amsmath}
\usepackage{bm}
\usepackage{cancel}
\usepackage{graphicx}

\usepackage{color}
\definecolor{headingcolour}{RGB}{50, 150, 190}

\usepackage{lipsum}

\usepackage[authoryear]{natbib}

\usepackage{setspace}

\usepackage{quattrocento}

\usepackage{sectsty}
\allsectionsfont{\fontfamily{phv}\selectfont}

\usepackage[T1]{fontenc} 

\usepackage[utf8]{inputenc}

\usepackage[margin=1in]{geometry}
\usepackage{fancyhdr}
\usepackage{caption}
\DeclareCaptionFormat{label}{\textbf{#1 #2} #3\hrulefill}
\usepackage[compact]{titlesec}

\usepackage{graphicx}

\usepackage{enumitem}
\setlist{nosep}

\usepackage[bookmarks=true]{hyperref}

\renewenvironment{abstract}{%
	\hspace{0.025\linewidth}\begin{minipage}{0.95\textwidth}
		\rule{\textwidth}{1pt}\small\selectfont}
	{\vspace{-0.5em}\par\noindent\rule{\textwidth}{1pt}\end{minipage}\vspace{1em}}

\makeatletter
\renewcommand{\maketitle}{\bgroup\setlength{\parindent}{0pt}
	\thispagestyle{empty}
	\begin{flushleft}
		{\bf \fontfamily{phv}\selectfont \LARGE \@title}
		
		\bf \fontfamily{phv}\selectfont \@author
	\end{flushleft}\egroup
}
\makeatother

\begin{document}
	\doublespacing
	\title{Modeling non-Fickian solute transport due to mass transfer and physical heterogeneity on arbitrary groundwater velocity fields}
	\author{Scott K. Hansen\footnote{Zuckerberg for Water Research, Ben-Gurion University of the Negev, Israel} and Brian Berkowitz\footnote{Department of Earth and Planetary Science, Weizmann Institute of Science, Israel}}
	\maketitle
	\singlespacing
	
	\begin{abstract}
		We present a hybrid approach to groundwater transport modeling, ``CTRW-on-a-streamline'', that allows continuous-time random walk (CTRW) particle tracking on large-scale, explicitly-delineated heterogeneous groundwater velocity fields. The combination of a non-Fickian transport model (in this case, the CTRW) with general heterogeneous velocity fields represents an advance of the current state of the art, in which non-Fickian transport models \textit{or} heterogeneous velocity fields are employed, but generally not both. We present a general method for doing this particle tracking that fully separates the model parameters characterizing macroscopic flow, subscale advective heterogeneity, and mobile-immobile mass transfer, such that each can be directly specified a priori from available data. The method is formalized and connections to classic CTRW and subordination approaches are made. Numerical corroboration is presented.
	\end{abstract}
	
	\section{Introduction}
	\subsection{Motivation}
		Classical Eulerian numerical models that employ a discretized advection-dispersion equation (ADE) are often not sufficient to model groundwater solute transport with needed realism because they fail to capture important physical heterogeneity and mass transfer at scales beneath their discretization scale. Additionally, their flow models may be oversmoothed even above their discretization scale due to a lack of constraining information. Non-Fickian transport models are a generalization of the classical transport models that can capture a wider range of physics, and have demonstrated success at capturing realistic chemical physics that cannot be modeled by classic approaches by introducing additional model flexibility. In the case of the popular continuous-time random walk (CTRW) approach, the additional flexibility comes in the free selection of a waiting time probability distribution, $\psi(t)$, that better describes the dispersion process.
				
		However, when capturing the effects of groundwater velocity heterogeneity, there has commonly been an ``all or nothing'' choice. Either model all the heterogeneity explicitly and then use an ADE-based transport model, or model all heterogeneity implicitly by considering its effects, combined with those of any other sources of non-Fickian behavior, in selection of the non-Fickian model parameters (e.g., $\psi(t)$). A consequence is that it is difficult to incorporate partial knowledge about large-scale flow into non-Fickian solute transport models, and it is generally difficult to predict correct model parameters a priori, as these depend in non-straightforward ways on the interaction of multiple physical and chemical sources of heterogeneity. This limitation may have reduced the penetration of advanced transport models from academic research into hydrogeologic practice. But whatever the underlying reasons, practitioners have typically favored explicitly heterogeneous models incorporating large-scale flow information coupled with ADE-based transport, whereas researchers have often chosen conversely: employing homogeneous, often quasi-1D, flow alongside more exotic non-Fickian transport models. Ideally, a hybrid approach to modeling would be available (a) allowing usage of explicitly-delineated, spatially-heterogeneous groundwater velocity fields, with non-Fickian transport modeling used \textit{only} to capture small-scale physics, and (b) whose non-Fickian model parameters have a straightforward relationship to the physics of the natural system they model. Developing such a hybrid approach is the goal of this work.
	
	\subsection{Sources of heterogeneity beneath flow model support scale}
	\label{sec: hetero below support}
	
		Before it developing a hybrid approach in which small-scale flow heterogeneity and mass transfer are captured with a stochastic transport model, it is naturally important to understand these processes and their impact on transport, and why a stochastic approach is valuable. 
	
		Consider an ideal scenario in which the groundwater flow field is known down to the pore scale in the plume region and computational power is unlimited, with the only stochastic process operative being molecular diffusion. In this situation, solute transport could in theory be captured exactly by a deterministic model, and perhaps by a discretized advection-dispersion equation. However, in realistic scenarios only the volume average groundwater velocity over some larger scale may be known with any precision, and the local-scale velocity fluctuations are random in an epistemic sense: unknown except potentially for their statistics. The first effort to deal with this problem was the so-called macrodispersion theory, which attempts encapsulate the effects of all the unknown velocity fluctuations into an additional Fickian dispersive term, justified by the Central Limit Theorem. While this approach is appropriate for large distances from the solute source and smaller heterogeneity \citep{Hansen2018}, frequently we are interested in understanding solute breakthrough in systems for which these restrictions are not appropriate. In such cases, non-Fickian behavior is generally observed and a more general approach such as usage of the continuous time random walk (CTRW)---where the travel time is captured by a travel time pdf---is appropriate.
		
		The groundwater velocity field is determined by solution of the groundwater flow equation on a (generally) heterogeneous hydraulic conductivity ($K$) field, and it is reasonable that its heterogeneity statistics will determine flow heterogeneity. On the assumption of lognormality of the $K$ field, (typically unrealistically) small log-$K$ variance, and multi-Gaussian correlation structure with light-tailed (exponentially decaying or finitely supported) semivariogram, the macrodispersion theory \citep{Rubin2003} provides analytical expressions for the solute transport behavior. However, when these assumptions are violated, we must rely on numerical studies linking heterogeneity and breakthrough curve statistics. There are not many of these in the literature, but to our knowledge, those that exist point strongly in two directions. For mild-to-moderate hydraulic conductivity heterogeneity and multi-Gaussian correlation structure, numerical studies have indicated a lognormal distribution for $\psi(t)$, gradually acquiring power law behavior in the tails as heterogeneity increases \citep{Gotovac2009,Hansen2018}. We stress that sub-Darcy-scale heterogeneity and trapping may still cause power law tailing even under mild or nonexistent $K$-field heterogeneity, as numerous experiments have shown (see discussion in \cite{Berkowitz2006}). For lognormally distributed $K$ and heavy-tailed (i.e., power law) semivariograms, studies have pointed in different directions. \cite{Moslehi2017} analyzed breakthrough in conductivity fields with a single, moderate, heterogeneity ($\sigma^2_{\ln K}$ = 4) and different strengths of power law semivariogram, finding their breakthrough curves generally well described by lognormal distributions, except for some heavy tailing. \cite{Bolster2012} considered mildly heterogeneous, but power law correlated, \textit{velocity} fields, showing analytically, under the validity conditions of their perturbation expansion, that breakthrough curves show power law tailing with the same exponent as the semivariogram. \cite{Zhang2013} performed a large-scale simulation study on highly heterogeneous realizations generated with depositional simulators (i.e., non-multivariate-Gaussian realizations), finding generally power-law behavior across realizations. They further found that where the islands of the lowest-$K$ material had a power law size distribution (loosely analogous to the correlation length in a semivariogram), that this determined the tail exponent. However, the relationship between exponents that they uncovered differed from that of \cite{Bolster2012}. \cite{Edery2014} performed a numerical particle tracking study on moderate-to-large heterogeneity ($\sigma^2_{\ln K}$ = 3 to 7) multivariate Gaussian fields, finding their breakthrough curves generally well described by a truncated power law distribution; we also speculate, based on analysis for \cite{Hansen2018}, that the lower-heterogeneity breakthrough curves would also be well described by lognormal distributions, again deviating towards power law behavior at late time. Finally, \cite{Tyukhova2016} considered an idealized problem of transport in a homogeneous medium with embedded spheres, each homogeneous and each with random (uniform) conductivity. They found that for truncated power law conductivity distribution, breakthrough curves were truncated power law, and that for (truncated) lognormal conductivity distributions, breakthrough curves were lognormal.
			
		In addition to the heterogeneous advection just discussed, a second category of transport process may also cause sub-support-scale heterogeneity: mobile-immobile mass transfer (MIMT), or broadly speaking, trapping processes. This category naturally includes adsorption, in which solute is truly immobilized, but also includes processes in which diffusion roughly orthogonal to the local advection direction moves solute into low-permeability regions in which advection is essentially inoperative. It is known that such physics may be captured by CTRW \citep{Berkowitz2006, Berkowitz2016}. However, they are also frequently treated by a variety of specialized modeling approaches, including retardation factors, explicit two-domain models \citep[e.g.,][]{Neretnieks1980}, dual porosity approaches \citep{Gerke1993}, kinetic sorption models \citep[see, e.g.,][]{Fetter1999}, multi-rate mass transfer (MRMT) \citep{Haggerty1995}, and fractal-MRMT / memory function approaches \citep{Schumer2003}. We note that heterogeneous advection has also been modeled with MRMT (the so-called MRMT-1 of \cite{Dentz2003}), but for our purposes we treat it separately.
		
	\subsection{Numerical modeling of transport with particle tracking}
		A common approach to numerical modeling of transport is to write the governing partial differential or integro-differential equation and then solve it analytically or semi-analytically using transform methods. However, this approach is generally only tenable in systems with spatially-uniform parameters. For systems with spatially non-uniform parameters, solution of the governing equation must be accomplished by spatial discretization. This introduces the problems of numerical dispersion and (in reactive models) numerical mixing. Where the governing equation contains a temporal integral, as in non-Fickian transport models, numerical instability becomes a concern, too. 
		
		Particle tracking approaches avoid these difficulties, and have been found to work well under very general conditions, as long as enough particles are used. A classic particle tracking algorithm typically uses a constant temporal user-specified step size, $\Delta_t$. The simplest possible particle update conditions apply in the case of pure advection (i.e., streamline tracing) iteration of each particle's position would be performed according to the following equations:
		\begin{linenomath*}
		\begin{eqnarray}
		\bm{x}_{n+1} &=& \bm{x}_n + \bm{v}(\bm{x}_n) \Delta_t \label{eq: pure adv}\\
		t_{n+1} &=& t_n + \Delta_t \label{eq: time update}
		\end{eqnarray}
		\end{linenomath*}
		To model Fickian dispersion one modifies \eqref{eq: pure adv} to read 
		\begin{linenomath*}\begin{equation}
		\bm{x}_{n+1} = \bm{x}_n + \bm{v}(\bm{x}_n) \Delta_t + \bm{\Delta}_{\bm{x},n}^D, \label{eq: with disp}
		\end{equation}\end{linenomath*}
		where in the case of isotropic diffusion
		\begin{linenomath*}\begin{equation}
			\Delta_{\bm{x},n}^D \equiv \bm{\eta},
			\label{eq: fickian disp}
		\end{equation}\end{linenomath*}
		where $\bm{\eta}$ is a random 3-vector, each of whose components $\eta_i \sim N(0,2 D \Delta_t)$, and $D$ is the Fick's Law constant. For CTRW particle tracking, either \eqref{eq: pure adv} or \eqref{eq: with disp} may be used, but the time step selection in \eqref{eq: time update} is modified so as to be selected at random:
		\begin{linenomath*}\begin{equation}
		\Delta_t \sim \psi(\Delta_t),
		\end{equation}\end{linenomath*}
		where $\psi$ is a probability distribution function with strictly non-negative support. 
		
		Below, we will see how to modify these equations to capture the sources of heterogeneity discussed in Section \ref{sec: hetero below support} while also incorporating coarse-scale flow information. Our strategy is to perform a CTRW that makes essentially fixed-length steps down the streamlines of the coarse-scale flow field. Such an approach is sometimes referred to as a time-domain random walk (TDRW), a special case of CTRW originally developed in 1D based on ADE transport solutions \citep{Banton1997,Reimus2002}, and subsequently developed to model MIMT in discrete fractures by \cite{Delay2001}. This approach was formalized using CTRW transition-time distribution theory by \cite{Cvetkovic2002} and explicit solutions for rock matrix residence times were presented by \cite{Painter2008}. Subsequently, the TDRW approach was employed in a large number of particle tracking studies on discrete fracture networks. For transport in heterogeneous porous media, \cite{Hansen2014} introduced a cognate CTRW approach, capturing transport as a series of transitions among parallel planes orthogonal to mean flow (leading to a quasi-1D upscaled transport model). Subsequently, another quasi-1D approach that approximates flow in heterogeneous media with an effective medium and also incorporates MIMT was discussed by \cite{Cvetkovic2016}. TDRW solutions based on classical (ADE) ideas and that include transverse dispersion were also presented by \cite{Bodin2015}. 
		
		Some other works whose concepts we build upon include \cite{Cortis2004}, who introduced a zoned approach to CTRW modeling of subsurface transport, with different CTRW parameters manually defined for different regions. The idea that a travel time distribution for advection with random motion can be expressed as the product of an average velocity and a distribution representing heterogeneity, is latent in the \cite{Kreft1978} solution for advective-dispersive breakthrough (see the factorization we present in \eqref{eq: KZ inv gauss}). This idea has been employed in the context of scaling CTRW transition time distributions to a local advection velocity by \cite{Srinivasan2010} and by \cite{Kang2014}. The concept of mapping an advective travel time to a random total time including multiple immobilization events was presented by \cite{Benson2009} for exponentially-distributed mobile and immobile times. A similar approach has been adopted for arbitrary immobile time distributions by \cite{Hansen2016} and by \cite{Russian2016}. Finally, the CTRW-on-a-streamline approach can be seen as extending the ideas of \cite{Cirpka2003} to general, potentially non-Fickian, transport: instead of using an additional dispersion term to capture small-scale flow field variability excluded from a smoothed deterministic model, we propose to use a CTRW transition time distribution.
		
		Developing from the ideas in these works, we present a complete treatment capturing local-scale heterogeneous advection, MIMT, and transverse dispersion using fully 3D streamlines and presenting explicit, physics-based formulae for the relevant CTRW transition-time distributions.
			 
	\section{Development of the CTRW-on-a-streamline approach}

	\subsection{Theory underlying the approach}
	\label{sec: theory}
		For pure streamline tracing, it is possible to choose a fixed \textit{distance}, $d$, that is traversed in equation \eqref{eq: pure adv} by adjusting the time step with reference to the local velocity: $\Delta_{t_O,n} = \frac{d}{\| \textbf{v}(\textbf{x}_n) \|}$. We add the subscript $n$ on each time step, $\Delta t_n$, because the time taken with each advective step is now variable and the variable $t_O$ stands for ``operational time'', whose significance will be explained below. This \textit{fixed distance increment, variable time} approach to large-scale advection is advantageous versus the common \textit{fixed time increment, variable distance} approach because it sets us up to use a CTRW transition time distribution to capture small-scale advective heterogeneity and mass transfer. To wit: we first \textit{define} a random walk transition to occur whenever another increment $d$ downgradient is traversed. We then seek to define the conditional pdf, $\psi(\Delta_{t_C}|\Delta_{t_O})$, for the true or \textit{clock} time, $\Delta_{t_C}$, taken to traverse that increment when small-scale physical heterogeneity and mass transfer are considered, conditioned on the notional or \textit{operational} time that would be taken to traverse the same segment by pure advection on the large-scale streamlines.
			
		Each segment of length $d$ of a large-scale streamline may be conceptualized as representing a flux-weighted ensemble, or ``bundle'', of small-scale, adjacent stream tubes of length $d$ with varying local harmonic mean hydraulic conductivity. Each small-scale stream tube has a different advection time, $\Delta_{t_A}$, and because the small-scale stream tubes in an ensemble are epistemically interchangeable (we have no knowledge of their exact position or of the particular small-scale tube to which a particle being tracked along a large-scale streamline belongs), it is sensible to define a conditional probability distribution, $\phi(\Delta_{t_A}|\Delta_{t_O})$.
			
		The nature of $\phi(\Delta_{t_A}|\Delta_{t_O})$ can be characterized by noting that the small-scale stream tubes in an ensemble share approximately the same gradient: justified by the fact that the global head field is an integral quantity and is smooth relative to the underlying hydraulic conductivity ($K$) field. Because $K$ is everywhere independent of the boundary conditions on head that determine the magnitude of the large-scale effective $\bm{v}$, and the head drop is shared by small-scale stream tubes in an ensemble, it follows that the Darcy velocities in each of the tubes in the ensemble always maintain a same constant of proportionality to $\|\bm{v}\|$, regardless of system boundary conditions. And because $d$ is fixed, the same is true of travel time, $\Delta_{t_A}$. Thus a simple scaling law applies: $\phi(\Delta_{t_A}|\Delta_{t_O})\Delta_{t_O}=\phi(\Delta_{t_A}/\Delta_{t_O}|1)=:f(r)$, where $f$ is some unknown pdf, solely determined by the statistics of small-scale heterogeneity, which contains all relevant information for mapping from operational time to true advection time, and $r=\Delta_{t_A}/\Delta_{t_O}$. Another way of looking at this is that the average of the ratio $\Delta_{t_A}$ to $\Delta_{t_O}$ is independent of the actual magnitude of $\Delta_{t_O}$ (and hence the magnitude of the groundwater velocity). The distribution $f(r)$ is determined by flow heterogeneity alone. 
		
		This informal argument is supported by a number of additional lines of evidence. We note that this is true for all advective-dispersive systems (see equation \eqref{eq: f ig}). The formal analysis of \cite{Tyukhova2016} considers transport in a regularly-spaced grid of equal-sized circular inclusions with random (low) conductivity in a uniform higher-conductivity matrix and shows mathematically that the velocity in any each inclusion varies in proportion with the system average flux. \cite{Comolli2016} show how under pure advection, the pdf for fixed distance travel time can be written in terms of the Lagrangian velocity pdf, and that under ergodic conditions the Eulerian velocity pdf determines the Lagrangian. Because the Eulerian pdf naturally scales linearly with global hydraulic gradient, the ratio of $\Delta_{t_A}$ to $\Delta_{t_O}$ must be independent of average velocity under ergodic conditions. 
			
		Once advective heterogeneity is specified by $f(r)$, any MIMT, whether due to diffusion into secondary porosity or to chemical adsorption, can be captured by defining two additional probability distributions. The first distribution is exponential, with a rate constant, $\lambda$, representing the probability of immobilization of mobile solute per unit time traveled. The second, not necessarily exponential, distribution, $g(t)$, represents the time for the length of a single particle immobilization event. The exponential form of the mobile-time distribution is justified by the assumption of randomly located sites that are dense relative to length scale $d$. We note that the $\lambda$ and $g$ approach captures even multiple types of immobilization sites with different capture and release distributions (i.e. multi-rate mass transfer), as the time until the first event for multiple simultaneous exponentially-distributed processes is itself exponential, and we can create an immobilization-site-prevalence-weighted average of the immobilization time pdf's for the various site types. 
		
		To quantify the travel time increase due to MIMT, we can develop a conditional pdf, $\zeta(\Delta_{t_C}|\Delta_{t_A})$. We note that the total (clock) time is the advection time, $\Delta_{t_A}$ plus the delay due to $n$ independent immobilization events each of whose lengths is drawn from $g(t)$. The probability mass function for $i$ immobilization events, $w(i)$, is Poisson distributed with parameter $\lambda \Delta_{t_A}$, leading \citep[][p. 262]{Billingsley1986} to probability mass function
		\begin{linenomath*}\begin{equation}
			w(i) = e^{-\lambda \Delta_{t_A}}\frac{(\lambda \Delta_{t_A})^i}{i!},
		\end{equation}\end{linenomath*} 
		Our conditional pdf, $\zeta$, is thus a weighted average of $i$-fold autoconvolutions of $g$:
		\begin{linenomath*}\begin{equation}
			\zeta(\Delta_{t_C}|\Delta_{t_A}) = \sum_{i=0} w(i)\times g^{(i*)}(\Delta_{t_C}-\Delta_{t_A}).
			\label{eq: zeta}
		\end{equation}\end{linenomath*}
		
		Where sites are sparse, an exponential distribution whose rate constant represents probability of immobilization of mobile solute per streamline \textit{distance} traveled may instead be more appropriate \citep{Margolin2003}. On this conception, we must understand $\lambda$ as capture probability per unit \textit{distance}, and $\zeta$ becomes no longer dependent on $\Delta_{t_A}$:  
		\begin{linenomath*}\begin{equation}
			w(i) = e^{-\lambda d}\frac{(\lambda d)^i}{i!},
		\end{equation}\end{linenomath*} 
		and we have
		\begin{linenomath*}\begin{equation}
			\zeta(\Delta_{t_C}|\Delta_{t_O},d) = \sum_{i=0} w(i)\times g^{(i*)}(\Delta_{t_C}-\Delta_{t_O}).
			\label{eq: zeta dist}
		\end{equation}\end{linenomath*}
		we will not discuss this second conception further, but it is important to stress its compatibility with our mathematics.
		
		Although it is most convenient computationally to work with $f$, $g$, and $\lambda$, we can easily relate the above analysis to the CTRW transition distribution, $\psi$, by constructing the conditional distribution 
		\begin{linenomath*}\begin{equation}
			\psi(\Delta_{t_C}|\Delta_{t_O}) = \int_0^{\Delta_{t_C}} \zeta(\Delta_{t_C}|\Delta_{t_A})\ \phi(\Delta_{t_A}|\Delta_{t_O})\ d\Delta_{t_A}.
		\end{equation}\end{linenomath*}
		This can be converted into a classic (unconditional) CTRW distribution by marginalizing
		\begin{linenomath*}\begin{equation}
			\psi(\Delta_{t_C}) = \int_0^{\Delta_{t_C}}\psi(\Delta_{t_C}|\Delta_{t_O})\ h_d(\Delta_{t_O}),
		\end{equation}\end{linenomath*}
		where $h_d$ is an unknown but determinate pdf for the operational time for all transitions of length $d$ anywhere in the domain. We see here two reasons why it is advantageous to employ the CTRW-on-a-streamline approach when large-scale information is available. First, the move $\psi(\Delta_{t_C}|\Delta_{t_O})$ to $\psi(\Delta_{t_C})$ involves a loss of information: under a classic approach, the same spatially-averaged $\psi(\Delta_{t_C})$ is used at all locations, inducing a loss of fidelity. Second, it is actually \textit{harder} to compute this distribution because it depends on an additional a priori unknown pdf that must be constrained by statistical information.
		
		Finally, we note that our analysis has so far detailed no mechanism for transfer of solute laterally between streamlines. This small-scale transverse dispersion is a Fickian process may be incorporated via the $\bm{\Delta}_{\bm{x},n}^D$ term in \eqref{eq: with disp}. Unlike the isotropic diffusion case \eqref{eq: fickian disp}, here we are only looking to model dispersion in the plane transverse to the streamline. First, we need to compute two orthonormal vectors which are also orthogonal to $\bm{v}$. To do so, an arbitrary vector $\textbf{a}$, not collinear with $\textbf{v}(\textbf{x})$, is specified. Then, orthonormal vectors are computed
		\begin{eqnarray}
			\bm{n}_1 &=& \frac{\bm{a} \times \bm{v}(\bm{x})}{\|\bm{a} \times \bm{v}(\bm{x}) \|_2} \label{eq: n1}\\
			\bm{n}_2 &=& \frac{\bm{n}_1 \times \bm{v}(\bm{x})}{\|\bm{n}_1 \times \bm{v}(\bm{x}) \|_2} \label{eq: n2}
		\end{eqnarray}
		For compactness, we define the matrix $\bm{N}(\bm{v})\equiv[\bm{n}_1\ \bm{n}_2]$ (i.e., the matrix whose columns are the streamline-orthogonal unit vectors $\bm{n}_1$ and $\bm{n}_2$), then we define $\bm{\eta}$ to be a 2-vector, each of whose components, $\eta_i \sim \mathcal{N}(0,2 \alpha_t d)$, where $\alpha_t$ represents the transverse dispersivity. Finally, the perturbation between streamlines is computed by matrix-vector multiplication:
		\begin{linenomath*}\begin{equation}
			\bm{\Delta}_{\bm{x},n}^D = \bm{N}(\bm{v})\bm{\eta}.
		\end{equation}\end{linenomath*}

	\subsection{Formal specification}
		It is straightforward to distill the above analysis into a set of formal equations that define the particle position and time updating rules for the CTRW-on-a-streamline approach. The top level equations are:
		\begin{eqnarray}
			\bm{x}_{n+1} &=& \bm{x}_n + \bm{v}(\bm{x}_n) \Delta_{t_O,n} + \bm{N}(\bm{v}(\bm{x}_n))\bm{\eta}, \label{eq: space update formal}\\
			t_{C,n+1} &=& t_{C,n} + \Delta_{t_C,n}, \label{eq: time update formal}
		\end{eqnarray}
		where
		\begin{eqnarray}
			\Delta_{t_O,n} &=& \frac{d}{\| \bm{v}(\bm{x}_n) \|}, \label{eq: operational from velocity} \\
			\eta_i &\sim& \mathcal{N}(0,2 \alpha_t d),\label{eq: dispersion formal}\\
			\Delta_{t_C,n} &\sim& \psi(\Delta_{t_C,n}|\Delta_{t_O,n}).\label{eq: clock update formal}
		\end{eqnarray}
		Placed in this form, it is clear that equations (\ref{eq: space update formal}-\ref{eq: clock update formal}) represent an evolution of the classic CTRW equations, with the conditioning of $\psi$ on $\Delta_{t_O,n}$ representing the major novelty (the restriction of transverse dispersion to the plane orthogonal to the streamline is a minor novelty). It is also clear from the examination of equations (\ref{eq: space update formal}-\ref{eq: time update formal}) that CTRW-on-a-streamline can be seen as a form of subordination technique \citep{Benson2009, Hansen2016}: particle location is updated according to a notional operational time, whereas particle time is updated in accordance with a true clock time.
	
	\section{Numerical implementation}
	\subsection{Pseudocode}
\begin{figure}
			\hspace{-5em}
			\includegraphics[scale=0.5]{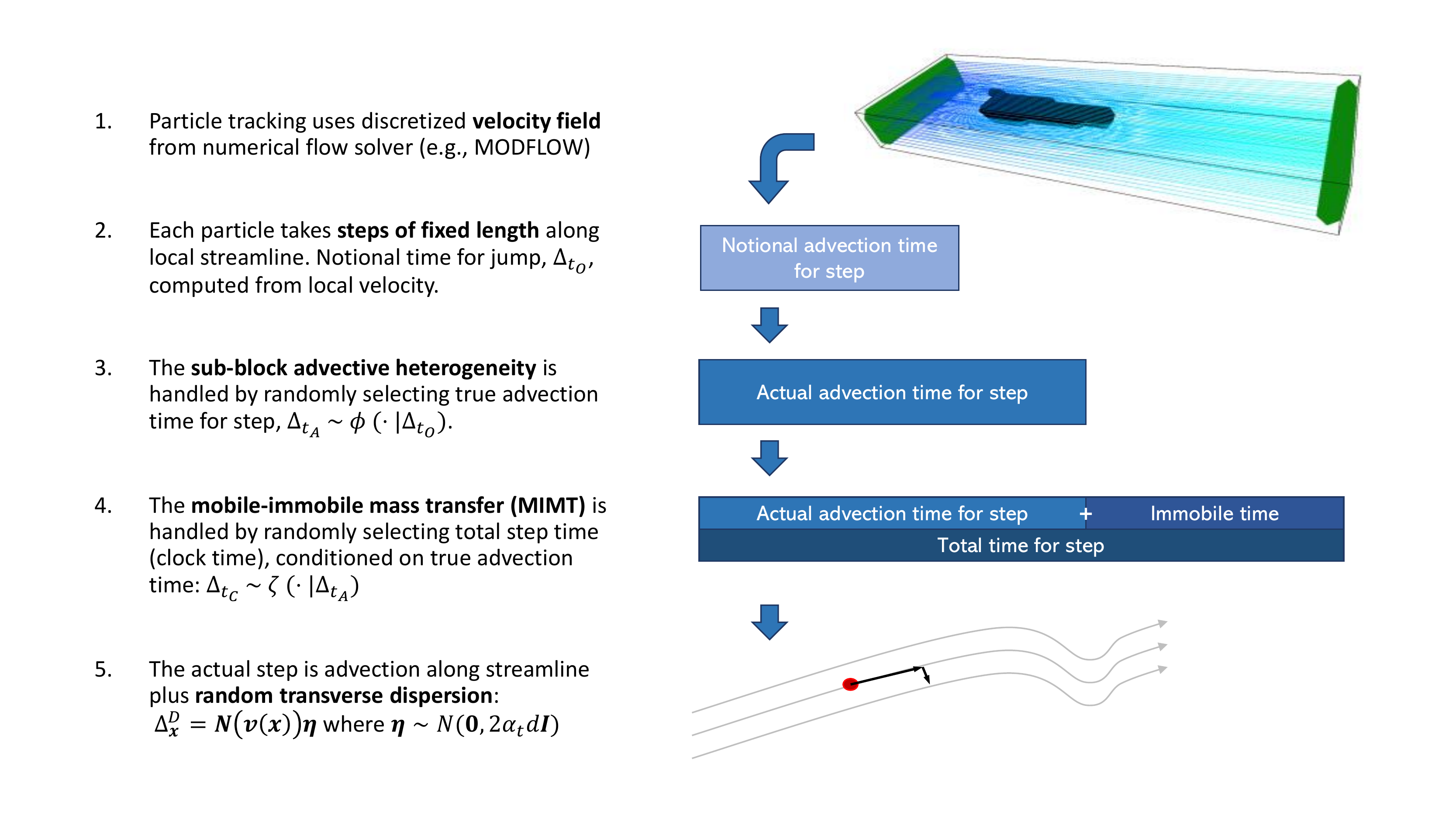}
			\caption{Schematic diagram outlining the essential conceptual steps of the CTRW-on-a-streamline approach.}
			\label{fig: schematic}
		\end{figure}

		As mentioned, it is not optimal to compute $\Delta_{t_C,n}$ by parameterizing the distribution $\psi(\Delta_{t_C,n}|\Delta_{t_O,n})$ and then drawing random variables from it. Rather, it is easier to build up $\Delta_{t_C,n}$ by first computing $\Delta_{t_A,n}$ by drawing from the pdf $f(\cdot)$, defining the advection time per unit operational time, and then computing the additional immobile time by making draws from $\mathrm{Poisson}(\cdot;\lambda \Delta_{t_A,n})$ and $g(\cdot)$. The following algorithm simulates a particle's transport in accordance with equations (\ref{eq: space update formal}-\ref{eq: clock update formal}), by means of these intermediate distributions:
		\begin{enumerate}
			\item For step $n$ = 0, initialize the particle's position in space and time by constructing the tuple $(\bm{x}_0,t_C)$ and decide on streamline tracing increment $d$.
			\item Referring to the explicit large-scale discrete velocity field, determine $\bm{v}(\bm{x}_n)$.
			\item Compute the operational time $\Delta_{t_O,n}$ by using \eqref{eq: operational from velocity}.
			\item Compute actual advection time $\Delta_{t_A,n}$ by drawing randomly from the advection-time-per-unit-operational-time pdf $f$, and multiplying that result by $\Delta_{t_O,n}$.
			\item Compute the number of immobilization events, $i$, in advection time $\Delta_{t_A,n}$ by drawing randomly from the distribution $\mathrm{Poisson}(\cdot;\lambda \Delta_{t_A,n})$.
			\item Draw $i$ random numbers independently from the $g(\cdot)$, the pdf for the length of a single sojourn in the immobile state, and sum them to yield the total time immobile time while traveling $d$ units down the streamline.
			\item Compute $\Delta_{t_C,n}$ by adding total immobile time and $\Delta_{t_A,n}$.
			\item Determine transverse dispersion by computing $\bm{N}(\bm{v}(\bm{x}_n))$ with (\ref{eq: n1}-\ref{eq: n2}), drawing the random vector, $\bm{\eta}$, according to \eqref{eq: dispersion formal}, and multiplying.
			\item Update the particle's position, $\bm{x}_{n+1}$, according to \eqref{eq: space update formal}.
			\item Update the particle's clock time, $t_{C,n+1}$, according to \eqref{eq: clock update formal}.
			\item Record any events of interest (breakthroughs or particle location snapshots).
			\item If $t_{C,n+1}$ is less than the end time for the simulation, increment the step index, $n$, by one and loop back to step 2.
		\end{enumerate}
		Note that because particles do not interact and have independent clocks, (a) this algorithm can be performed in parallel for multiple particles, and (b) particles can be initialized at different clock times. A schematic diagram summarizing the essential procedural steps is given in Figure \ref{fig: schematic}.
		
		Of course, to completely specify the algorithm, we need to specify $f(\cdot)$, $\lambda$, and $g(\cdot)$. How this is done depends on the underlying physics to be modeled; considerations are discussed immediately below.
	
	\subsection{Specification of advective heterogeneity}
		Advective heterogeneity refers to varying local pore water velocities experienced by particles as they are advected along a stream tube and experience transverse diffusion. For our purposes, it also incorporates physics that are sometimes considered using MRMT techniques (i.e., the ``MRMT-1'' class outlined in \cite{Dentz2003}), namely advection into discrete low-velocity features. The CTRW-on-a-streamline approach captures advective heterogeneity by directly encoding the spectrum of times taken to travel $d$ in the small-scale stream tubes associated with a large-scale streamline in the distribution $\phi(\Delta_{t_A}|\Delta_{t_O})$, or equivalently in $f(\cdot)$. For predictive modeling purposes, we would like to express the parameters defining the distribution $f(r)$ in terms of well-known parameters describing the subsurface heterogeneity, as in the case of MIMT. For the case of homogeneous local-scale Fickian transport, this is easily accomplished analytically: an inverse Gaussian distribution applies. However, heterogeneity at all scales complicates the picture. 
		
		Although, as outlined in the introduction, there is no general theory relating ensemble breakthrough parameters to conductivity field statistics, there are three classes of pdf supported by the literature for capturing advective heterogeneity pdf $f(r)$, each suitable for different problems. These classes are: (i) inverse Gaussian, which derives from Fickian analysis, (ii) lognormal, applicable to mildly or moderately heterogeneous systems, and (iii) power law / truncated power law, perhaps the most widely applicable, capturing the strong heterogeneity often encountered at field scale. We discuss how to specify each for use with the CTRW-on-a-streamline approach in turn.
		
		\subsubsection{Fickian dispersion: inverse Gaussian behavior}
			For longitudinal, local-scale Fickian dispersion, the travel time pdf for a transition of length $d$, is simply the well-known Inverse Gaussian distribution, as calculated by, e.g., \cite{Kreft1978}. We write it in factored form as
			\begin{linenomath*}\begin{equation}
				\phi(\Delta_{t_A}) = \frac{1}{\Delta_{t_A}} \left[\frac{d}{\sqrt{4\pi\alpha_l\|v\|\Delta_{t_A}}}\exp\left\{ \frac{-(d - \|v\|\Delta_{t_A})^2}{4\alpha_l\|v\|\Delta_{t_A}} \right\}\right],
				\label{eq: KZ inv gauss}
			\end{equation}\end{linenomath*}
			where, $\alpha_l$ is the Fickian longitudinal dispersivity. Note that the quantity in square brackets is dimensionless. Further noting that $r = \Delta_{t_A}/\Delta_{t_O} = \|v\|\Delta_{t_A}/d$, we may rephrase in terms of $r$:
			\begin{linenomath*}\begin{equation}
				f(r)=\frac{1}{r\sqrt{4\pi A r}}\exp\left\{-\frac{(r-1)^2}{4Ar}\right\},
				\label{eq: f ig}
			\end{equation}\end{linenomath*}
			where we also define the dimensionless variable $A\equiv\alpha_l/d$. 
			
			In Section \ref{sec: theory}, we present a number of strands of evidence supporting the contention that the distribution of the ratio $\Delta_{t_A}/\Delta_{t_0}$ is independent of the large-scale groundwater velocity, and is rather a proxy for small-scale heterogeneity. The analysis deriving \eqref{eq: f ig}, which does not depend explicitly on either $\Delta_{t_A}$ or $\Delta_{t_O}$ from \eqref{eq: KZ inv gauss}, shows that this is true for all systems described by the advection-dispersion equation. 
			 
		\subsubsection{Lognormal behavior}
			The lognormal travel time distribution is defined by the variance of log arrival time distributions, $\sigma_{\ln t}^2$, rather than $\alpha_l$, but is otherwise similar. For moderate levels of heterogeneity, $\sigma_{\ln t}^2$ can be predicted from $\sigma_{\ln K}^2$ and correlation length by  numerical study \citep{Gotovac2009,Beaudoin2013,Hansen2018}. Provided that $\Delta_{t_O}$ is the geometric mean of the travel times in its corresponding stream tube bundle, we may write exactly:
			\begin{linenomath*}\begin{equation}
				f(r)=\frac{1}{r\sqrt{2\pi\sigma_{\ln t}^2}} \exp\left\{-\frac{(\ln r+\frac{1}{2}\sigma_{\ln t}^2)^2}{2\sigma_{\ln t}^2}\right\}.
				\label{eq: f ln}
			\end{equation}\end{linenomath*}
			The assumption that large-scale streamline velocity \textit{is} the geometric mean of the corresponding small-scale stream tube velocities restates the assumption that the effective hydraulic conductivity for a region is its geometric mean conductivity \citep{Tsang1994} (discussed at length in chapter 5 of \cite{Rubin2003}), combined with our general assumption that the head field being smooth relative to the conductivity field.  
			
		\subsubsection{Power law behavior}
			It is straightforward to write a transition time pdf for a pure power law distribution (i.e. Lomax distribution):
			\begin{linenomath*}\begin{equation}
				f(r)=\frac{\alpha\lambda}{(r+\lambda)^{1+\alpha}}.
				\label{eq: f pl}
			\end{equation}\end{linenomath*}
			General theory relating the parameters $\alpha$ and $\lambda$ to aquifer heterogeneity statistics does not appear to exist in the literature, unlike for Fickian and lognormal breakthrough, although relations have been developed \citep[e.g.][]{Edery2014,Nissan2019}.
	
	\subsection{Specification of MIMT}
	\label{sec: MIMT}
		The formulation of MIMT is that first employed by \cite{Margolin2003}, corresponding to an Eulerian transport equation of the form
		\begin{linenomath*}\begin{equation}
			\frac{\partial c_\mathrm{m}}{\partial t} + \lambda c_\mathrm{m} - \int_{0}^{t} g(t-\tau) \lambda c_\mathrm{m}(\tau) d\tau = L\left\{c_\mathrm{m}\right\},
			\label{eq: our MIMT}
		\end{equation}\end{linenomath*}
		where $L$ is a linear transport operator, and we use $c_\mathrm{m}$ to refer explicitly to the mobile concentration. This differs superficially from many of the approaches to MIMT in the literature, so it is necessary to outline how this relates to them, and how $\lambda$ and $g(t)$, introduced in Section \ref{sec: theory}, may be determined explicitly from the parameters defining other MIMT formulations seen in the literature. We consider three such cases in the next subsections.
	
		\subsubsection{Connection to MRMT memory function}
			MRMT is a formulation that obviously generalizes single-rate mass transfer, but also dual-porosity models, explicit models of diffusion into secondary structures such as slabs, spheres, and cylinders, and slow kinetic sorption. The general MRMT transport equation has the form
			\begin{linenomath*}\begin{equation}
				\frac{\partial c_\mathrm{m}}{\partial t} + \frac{\partial c_\mathrm{im}}{\partial t}   = L\left\{c\right\},
				\label{eq: normal MIMT}
			\end{equation}\end{linenomath*}
			and needs to be closed by means of a relationship between $c_\mathrm{im}$ and $c_\mathrm{m}$. This relationship is most commonly expressed as a convolution,
			\begin{linenomath*}\begin{equation}
				\frac{\partial c_\mathrm{im}}{\partial t} = G*\frac{\partial c_\mathrm{m}}{\partial t},
				\label{eq: standard im-m relation}
			\end{equation}\end{linenomath*}
			where $G(\cdot)$ is a ``memory function''. This is the formulation used, for example, in the papers of \cite{Carrera1998} and \cite{Haggerty2000} (although a typo in the first reference's equation 13 causes it to be missing the time derivative on its LHS), and also in the fractal MIM formulation of \cite{Schumer2003}. Similarly to \cite{Haggerty2000}, we may integrate by parts to yield
			\begin{linenomath*}\begin{equation}
				\frac{\partial c_\mathrm{im}}{\partial t} = G(0)c_\mathrm{m}(t) - \cancelto{0}{G(t)c_\mathrm{m}(0)} + \frac{\partial G}{\partial t}*c_\mathrm{m}.
				\label{eq: standard im-m final}
			\end{equation}\end{linenomath*}
			The canceled term arises from the fact that \eqref{eq: standard im-m relation} is derived for an initially zero-concentration domain. (See equation 5 of \cite{Schumer2003} for the equivalent version of \eqref{eq: standard im-m relation} with a general initial condition.) An apparently different formulation relating the immobile concentration to the mobile concentration is used by \cite{Dentz2003}:
			\begin{linenomath*}\begin{equation}
				c_\mathrm{im} = \int_0^t G(t-\tau)c_\mathrm{m}(\tau)d\tau.
				\label{eq: dentz im-m relation}
			\end{equation}\end{linenomath*}
			However, when differentiated with respect to $t$, it yields
			\begin{linenomath*}\begin{equation}
				\frac{\partial c_\mathrm{im}}{\partial t} = G(0)c_\mathrm{m}(\tau) + \frac{\partial G}{\partial t}*c_\mathrm{m},
				\label{eq: dentz im-m final}
			\end{equation}\end{linenomath*}
			which is consistent with \eqref{eq: standard im-m final}. From inspection of either \eqref{eq: standard im-m final} or \eqref{eq: dentz im-m final}, it is obvious that the definitions
			\begin{eqnarray}
				\lambda &\equiv& G(0), 
				\label{eq: MRMT-CTRW equivalence lambda}\\
				g(t) &\equiv& \frac{\partial G}{\partial t},
				\label{eq: MRMT-CTRW equivalence g}
			\end{eqnarray}
			make \eqref{eq: our MIMT} equivalent to \eqref{eq: normal MIMT} combined with \eqref{eq: standard im-m relation}. 
			
			The careful reader may note that \cite{Dentz2003} \textit{also} showed how a CTRW memory function could be derived that is equivalent to any MRMT model, and that the corresponding CTRW transition time pdf is different from our $\zeta$, when defined according to \eqref{eq: zeta} and (\ref{eq: MRMT-CTRW equivalence lambda}-\ref{eq: MRMT-CTRW equivalence g}). The reason for this is that the earlier authors used a different definition of ``immobile'' from that introduced by \cite{Margolin2003} and used in this paper. \cite{{Dentz2003}} work by upscaling a microscopic (discrete site) formulation of the CTRW and consequently \textit{define} immobile particles to be those that are, at any instant, not completing transitions. However, in this paper, mesoscopic transitions are defined by periodic arrivals at ``milestones'' separated by distance $d$ on a streamline. Thus, we must consider mobile particles that are at a given moment undergoing advection but are not presently arriving at a milestone and completing a transition. In our conception, the particles that are immobilized due to sorption or diffusion into secondary porosity are a subset of the particles not presently completing transitions. This difference in conceptions accounts for the difference in equations.
			
		\subsubsection{Connection to first-order mass transfer coefficient}
			First-order mass transfer is described by the governing equation
			\begin{linenomath*}\begin{equation}
				\frac{\partial c_\mathrm{m}}{\partial t} + \mu (c_\mathrm{im}-c_\mathrm{m})  = L\left\{c_\mathrm{m}\right\}.
				\label{eq: first-order MIMT}
			\end{equation}\end{linenomath*}
			While this is plainly a special case of single rate mass transfer with equal rate constants for both mobilization and immobilization, and this is itself a subset of MRMT, there is no memory function in this formulation, and so separate analysis is required to derive $\lambda$ and $g(t)$. We note that if we define $g(t)=\mu \exp(-\mu t)$, then the probability of a particle remaining immobile for at least $T$ is $\int_T^\infty \mu \exp(-\mu t) dt = \exp(-\mu T)$. The immobile concentration at time $t$ is the integral of the mobile-immobile flux at each previous time, $\tau$, weighted by the probability of remaining immobile for at least $T=t-\tau$. Thus, it follows that
			\begin{eqnarray}
				c_\mathrm{im} &=& \int_{0}^{t} e^{-\mu(t-\tau)} [-\lambda c_\mathrm{m}(\tau)] d\tau,\\
				-\mu c_\mathrm{im} &=& \int_{0}^{t} g(t-\tau) \lambda c_\mathrm{m}(\tau) d\tau.
			\end{eqnarray}
			It then follows from direct inspection that \eqref{eq: our MIMT} and \eqref{eq: first-order MIMT} are equivalent, as long as the following identification is made:
			\begin{eqnarray}
				\lambda &\equiv& \mu,
				\label{eq: FO-CTRW equivalence lambda}\\
				g(t) &\equiv& \mu e^{-\mu t}.
				\label{eq: FO-CTRW equivalence g}
			\end{eqnarray}
		
		\subsubsection{Connection to retardation factor}
			The final case we consider is a transport equation where MIMT is encoded by multiplication of the time derivative by a retardation factor, R:
			\begin{linenomath*}\begin{equation}
			R\frac{\partial c_\mathrm{m}}{\partial t} = L\left\{c_\mathrm{m}\right\}.
			\label{eq: retardation MIMT}
			\end{equation}\end{linenomath*}
			We note that this model is an inherently defective simplification of first-order mass transfer. The defect lies in the fact that MIMT always causes dispersion which is erroneously disregarded by use of the retardation factor alone. The so-called ``local equilibrium assumption'' actually amounting to the assumption that dispersion due to MIMT is small relative to other sources of dispersion \citep{Hansen2018a}. Thus, our approach is to begin with a single-rate mass transfer model, and define its parameters so that its effective velocity corresponds to that of the retardation model, and its dispersion due to mass transfer is ``small'' in some sense. This can not be said to have less realism than \eqref{eq: retardation MIMT}, and can generate arbitrarily close behavior. We begin from a slightly generalized form of \eqref{eq: first-order MIMT}, with distinct mobilization and immobilization rates:
			\begin{linenomath*}\begin{equation}
			\frac{\partial c_\mathrm{m}}{\partial t} + \mu c_\mathrm{im} - \lambda c_\mathrm{m}  = L\left\{c_\mathrm{m}\right\}.
			\label{eq: first-order MIMT relaxed}
			\end{equation}\end{linenomath*}
			It is easy to show \citep[e.g.,][]{Hansen2018a} that an equivalent $R$ for any $\lambda$ and $\mu$ is defined
			\begin{linenomath*}\begin{equation}
				R = 1 + \frac{\lambda}{\mu}.
			\end{equation}\end{linenomath*}
			It has been similarly been shown by \cite{Michalak2000} and \cite{Uffink2012} that the effective late-time Fickian dispersion coefficient, $D_\mathrm{eff}$, corresponding to the MIMT in \eqref{eq: first-order MIMT relaxed} is 
			\begin{linenomath*}\begin{equation}
				D_\mathrm{eff} = \frac{\lambda\mu v^2}{(\lambda+\mu)^3},
			\end{equation}\end{linenomath*}
			where $v$ is the magnitude of the streamline advection velocity. Thus, for \eqref{eq: our MIMT} to match \eqref{eq: retardation MIMT}, we must select $\lambda$ and $\mu$ in \eqref{eq: first-order MIMT relaxed} to match $R$, and so that $D_\mathrm{eff}$ is sufficiently small. This is attainable because $R$ fixes only the ratio of $\lambda$ and $\mu$, and their respective magnitudes can be scaled by any factor, $k$, to scale $D_\mathrm{eff}$ by the factor $k^{-1}$. We can bring \eqref{eq: our MIMT} and \eqref{eq: retardation MIMT} into arbitrary close alignment by choosing sufficiently large $\mu$ and making the following identification:
			\begin{eqnarray}
				\lambda &\equiv& (R-1)\mu,
				\label{eq: retardation-CTRW equivalence lambda}\\
				g(t) &\equiv& \mu e^{-\mu t}.
				\label{eq: retardation-CTRW equivalence g}
			\end{eqnarray}

	\section{Numerical corroboration}
		While the arguments connecting common MIMT models to the $\lambda$ and $g(t)$ used for CTRW-on-a-streamline modeling in Section \ref{sec: MIMT} are mathematically exact and do not need further support, the arguments supporting use of the single pdf, $f(r)$, defined by one of the equations (\ref{eq: f ig}-\ref{eq: f pl}) to capture the effects of small scale heterogeneous advection are more qualitative. As a result it is useful to demonstrate their successful use numerically. We shall do so, simultaneously demonstrating the entire CTRW-on-a-streamline procedure.
		
		Our approach is to generate a single ``true'' multi-Gaussian log hydraulic conductivity field and discretize it on a structured grid, generating a matrix of log hydraulic conductivity values. By convolution smoothing of this master field, a derivative field with less small-scale detail but sharing the same large-scale features may be created. By solving the steady-state groundwater flow equation on both conductivity fields with the same boundary conditions, discretized velocity fields may be created for each, and CTRW-on-a-streamline particle tracking performed on each. We seek to show that we recover concentration profiles and breakthrough curves that closely match those computed using the true velocity field by using the smoothed velocity field as well as an appropriate $f(r)$.
		
\begin{figure}
			\hspace{-5em}
			\begin{tabular}{>{\centering\arraybackslash} m{0.5cm} >{\centering\arraybackslash} m{8cm} >{\centering\arraybackslash} m{8cm}}
				&\textbf{True fields} & \textbf{Smoothed fields} \\
				\rotatebox{90}{$\bm{\log K}$ \textbf{fields}}&
				\includegraphics[trim={2cm 0 2cm 0.5cm}, clip, scale=0.65]{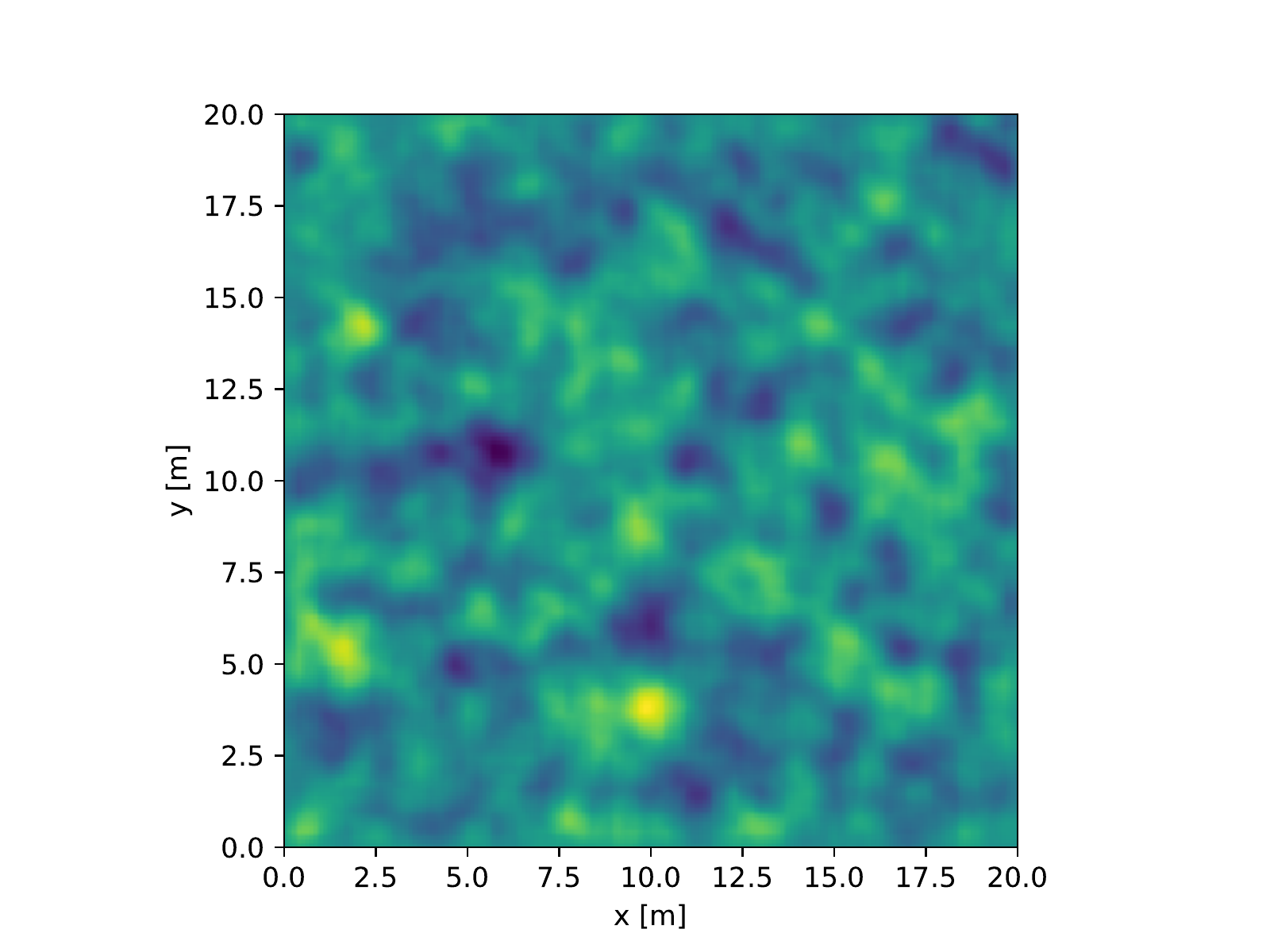} &
				\includegraphics[trim={2cm 0 2cm 0.5cm}, clip, scale=0.65]{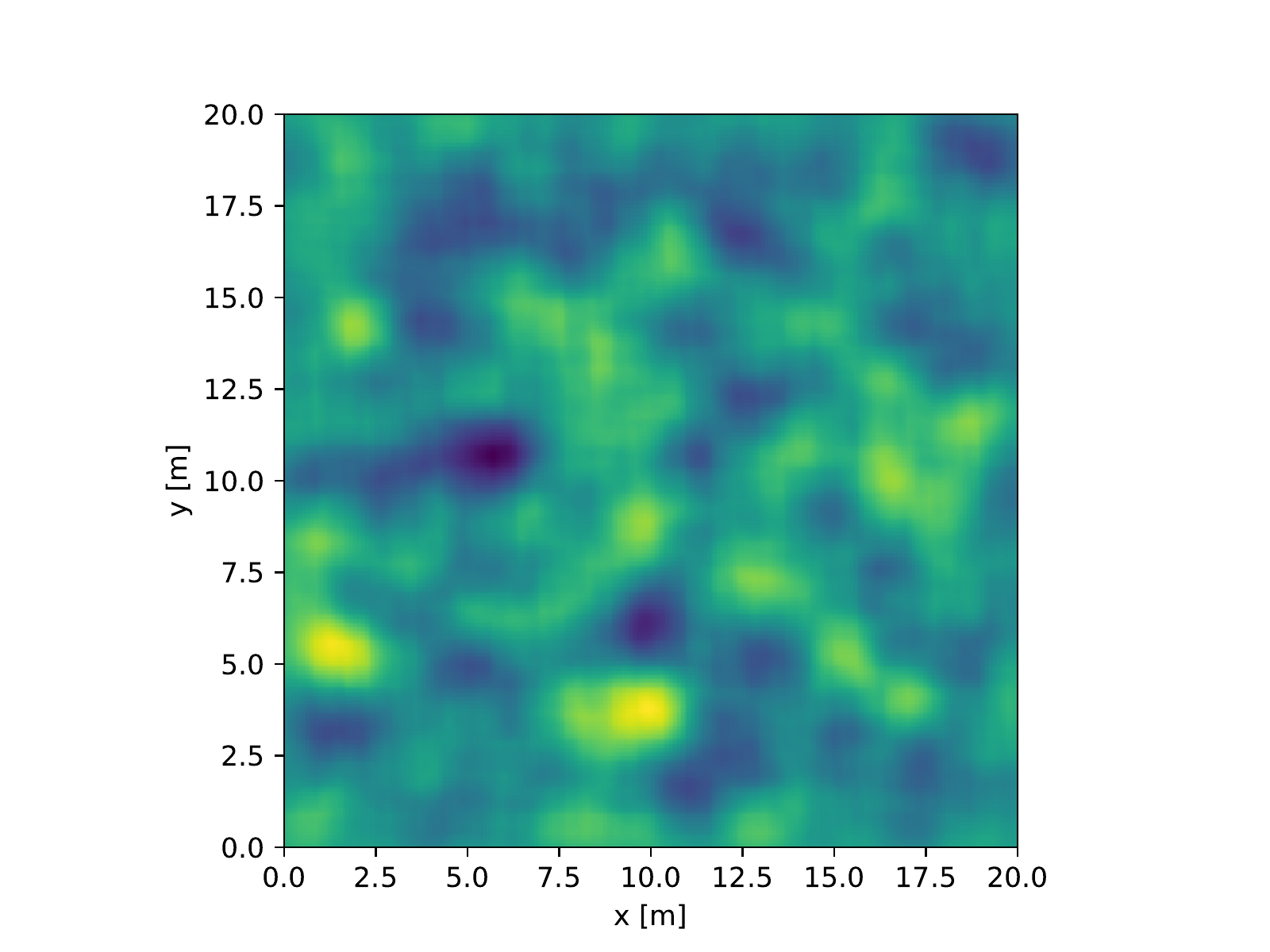} \\
				\rotatebox{90}{\textbf{Head and velocity fields}}&
				\includegraphics[trim={2cm 0 2cm 0.5cm}, clip, scale=0.65]{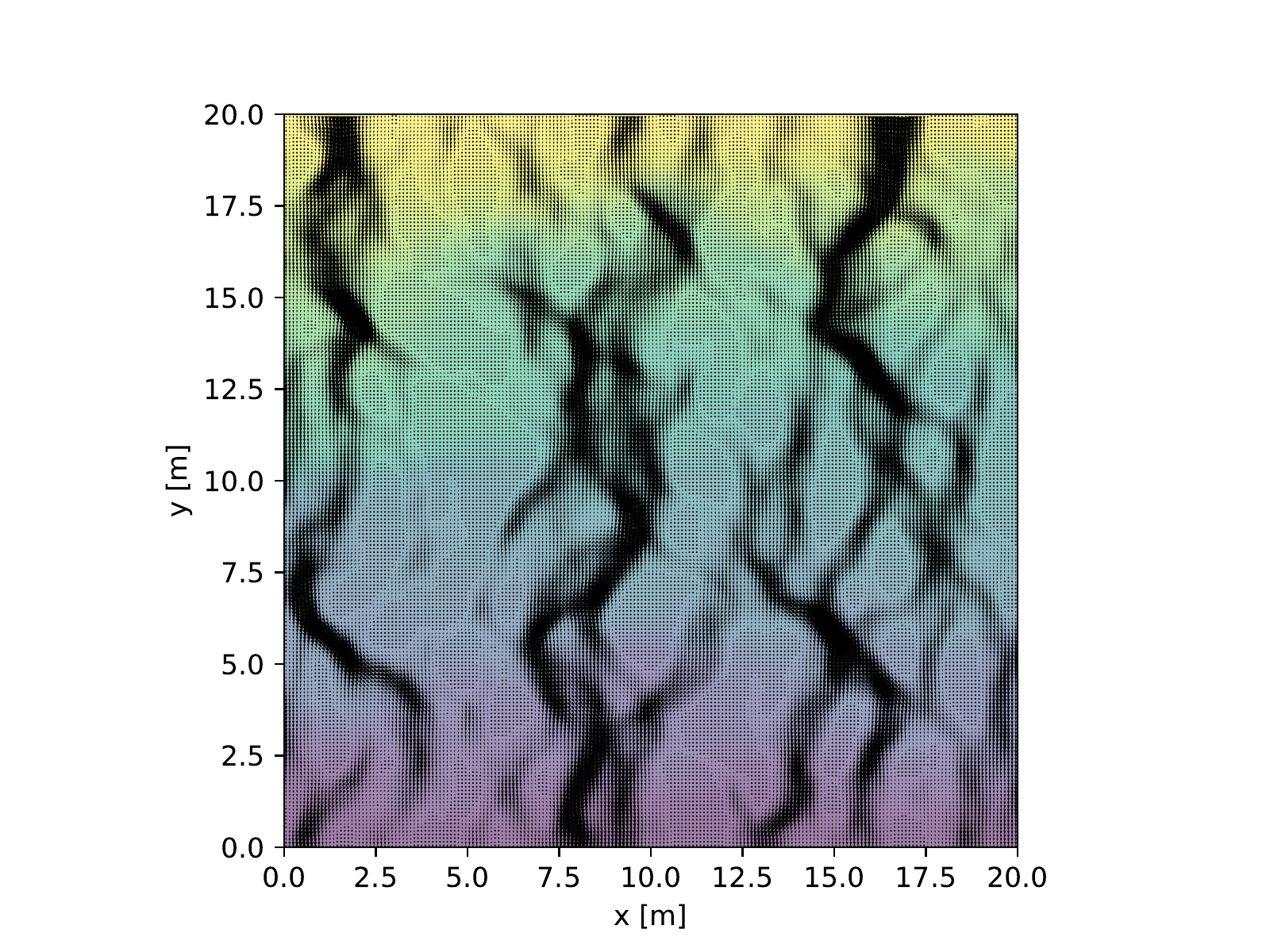} &
				\includegraphics[trim={2cm 0 2cm 0.5cm}, clip, scale=0.65]{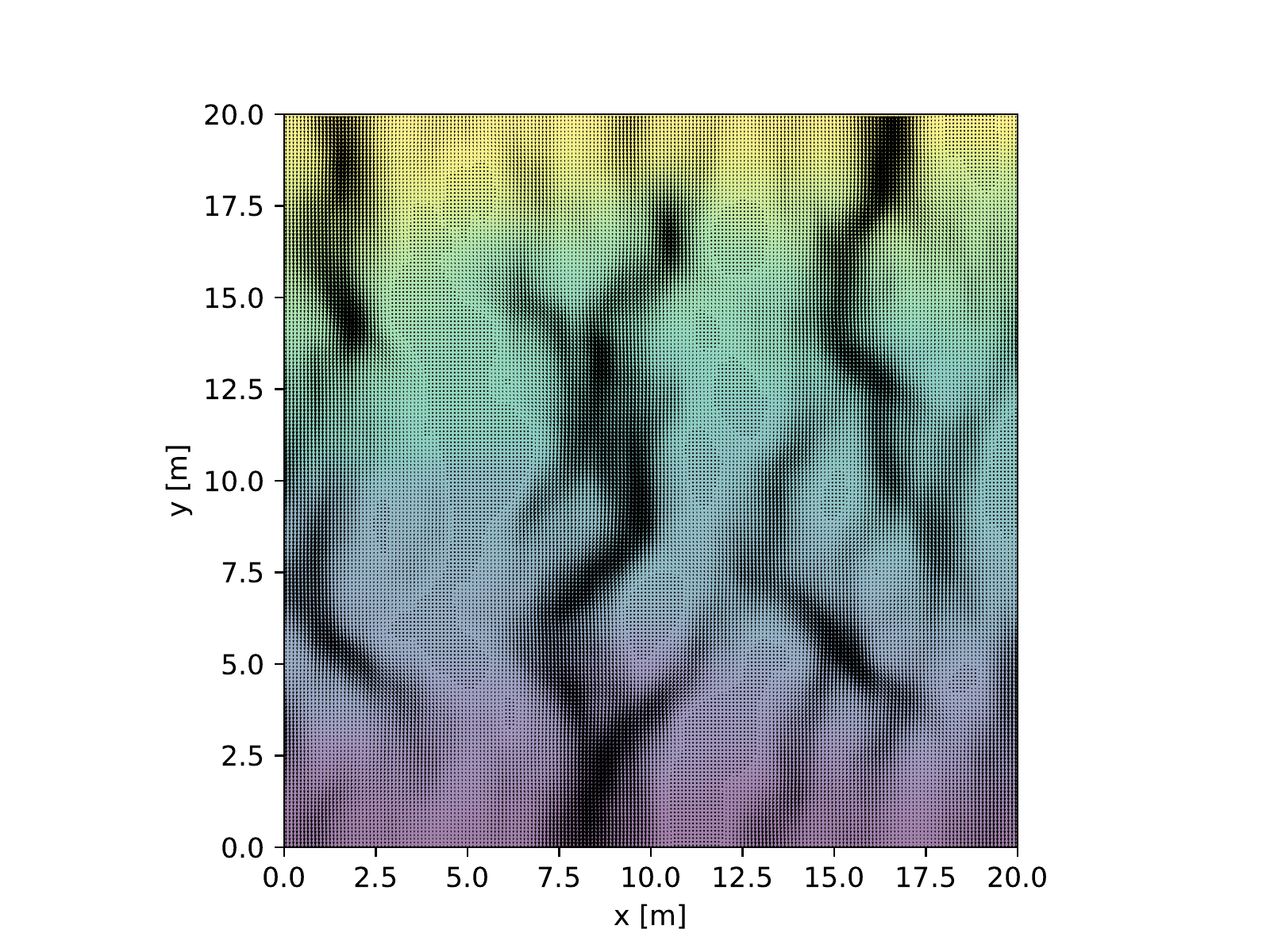}
			\end{tabular}
			\caption{Illustration of log-hydraulic hydraulic conductivity, head, and velocity fields used in the simulations. In the first column are the spatially detailed fields treated as true. In the second column are spatially smoothed case fields that approximate the true fields in the first column. In the top row, log conductivity fields are shown, with lighter color indicating higher conductivity. In the bottom row, the corresponding hydraulic head fields are shown, with lighter color indicating greater head. Superimposed on each head field is a black quiver plot in which each arrow's magnitude indicates the cell-center velocity for each cell in the discretization. (Because cell resolution is high, each arrow is quite small.)}
			\label{fig: k and h comparison}
		\end{figure}

		In more detail: a multivariate 20 m square master hydraulic conductivity field was generated using the GSTools package \citep{Muller2019} with an 0.1 m discretization length and geometric mean conductivity of $10^{-2}\  \mathrm{ms^{-1}}$. The (base 10) log hydrualic conductivity featured a multivariate Gaussian correlation structure with $\sigma^2_{\log K}$ = 0.5 and an exponential semivariogram with correlation length 0.5 m. From the 200 by 200 matrix representing the master conductivity field, the true conductivity field was derived by convolving against a uniform kernel with 0.8 m square support (represented by an 8 by 8 matrix). The smoothed conductivity field was derived by convolving against a uniform kernel with 1.6 m square support (represented by a 16 by 16 matrix). Both of the derived $\log K$ fields are shown in Figure \ref{fig: k and h comparison}. Both matrix convolutions, and all of the numerical analysis was performed in Python using the Numpy/Scipy ecosystem \citep{Oliphant2007}.
		
		For each of the true and smoothed conductivity fields, a head field was derived by numerically solving the steady-state governing equation
		\begin{linenomath*}\begin{equation}
			\nabla \cdot \left( K(x,y) \nabla h(x,y) \right) = 0,
		\end{equation}\end{linenomath*}
		subject to boundary conditions
		\begin{eqnarray}
			h(x, 0) &=& 0, \\
			h(x, 20) &=& 1, \\
			\frac{\partial h}{\partial x}(0, y) &=& 0, \\
			\frac{\partial h}{\partial x}(20, y) &=& 0,
		\end{eqnarray}
		where $h(x,y)\ [\mathrm{m}]$ represents the hydraulic head at coordinate $(x,y)$. Solution was computed using second-order accurate finite difference techniques, employing the Scipy sparse matrix solver tools. The resulting head distributions are shown in Figure \ref{fig: k and h comparison}, and the relative smoothness of the $h$ field relative to the $\log K$ field, as posited when developing the CTRW-on-a-streamline approach, is plainly apparent. 
		
		From each head field, a velocity field, $\bm{v}(x,y)$ was determined by solving 
		\begin{linenomath*}\begin{equation}
			\bm{v}(x,y) = \frac{K(x,y) \nabla h(x,y)}{\theta},
		\end{equation}\end{linenomath*}
		where $\theta=0.25$ was taken to be the spatially-uniform porosity, and cell center velocities were stored for each cell. These velocities are displayed as a quiver plot, again in Figure \ref{fig: k and h comparison}.
		
		Finally, three different particle tracking simulations were performed, each by flux-weighted injection of 10000 particles along the top ($y=20$) boundary between $x=1$ and $x=19$. This was implemented by assigning weights to each cell adjacent to the relevant portion of the $y=20$ boundary according to the magnitude of the $y$-component of $\bm{v}$ at their respective cell centers. Subsequently, a weight-proportional portion of the 10000 particles was assigned to each of these border cells and these particles were initially positioned uniformly randomly within said cell. For each of the three simulations, particle tracking was performed using the CTRW-on-a-streamline approach and two events were recorded: (a) the location of all particles when $t = 7\times 10^3$ s, and (b) the time of breakthrough of each particle at $y=0$. The three simulations were:
		 \vspace{-0.6\topsep}
		\begin{enumerate}
			\item Classical physics on the highly resolved field: $f(t)=\delta(t)$, $\alpha_t = 0.01$ m, with no MIMT.
			\item Classical physics on the smoothed field: $f(t)=\delta(t)$, $\alpha_t = 0.01$ m, with no MIMT.
			\item Additional CTRW physics on the smoothed field: $f(t)$ inverse Gaussian with $\alpha_l=0.152$ m, $\alpha_t = 0.01$ m, with no MIMT.
		\end{enumerate}
	
\begin{figure}
			\hspace{-5em}
			\begin{tabular}{>{\centering\arraybackslash} m{0.5cm} >{\centering\arraybackslash} m{8cm} >{\centering\arraybackslash} m{8cm}}
				&\textbf{On true velocity field} & \textbf{On smoothed velocity field} \\
				\rotatebox{90}{\textbf{With classical physics only}}&
				\includegraphics[trim={2cm 0 2cm 0.5cm}, clip, scale=0.65]{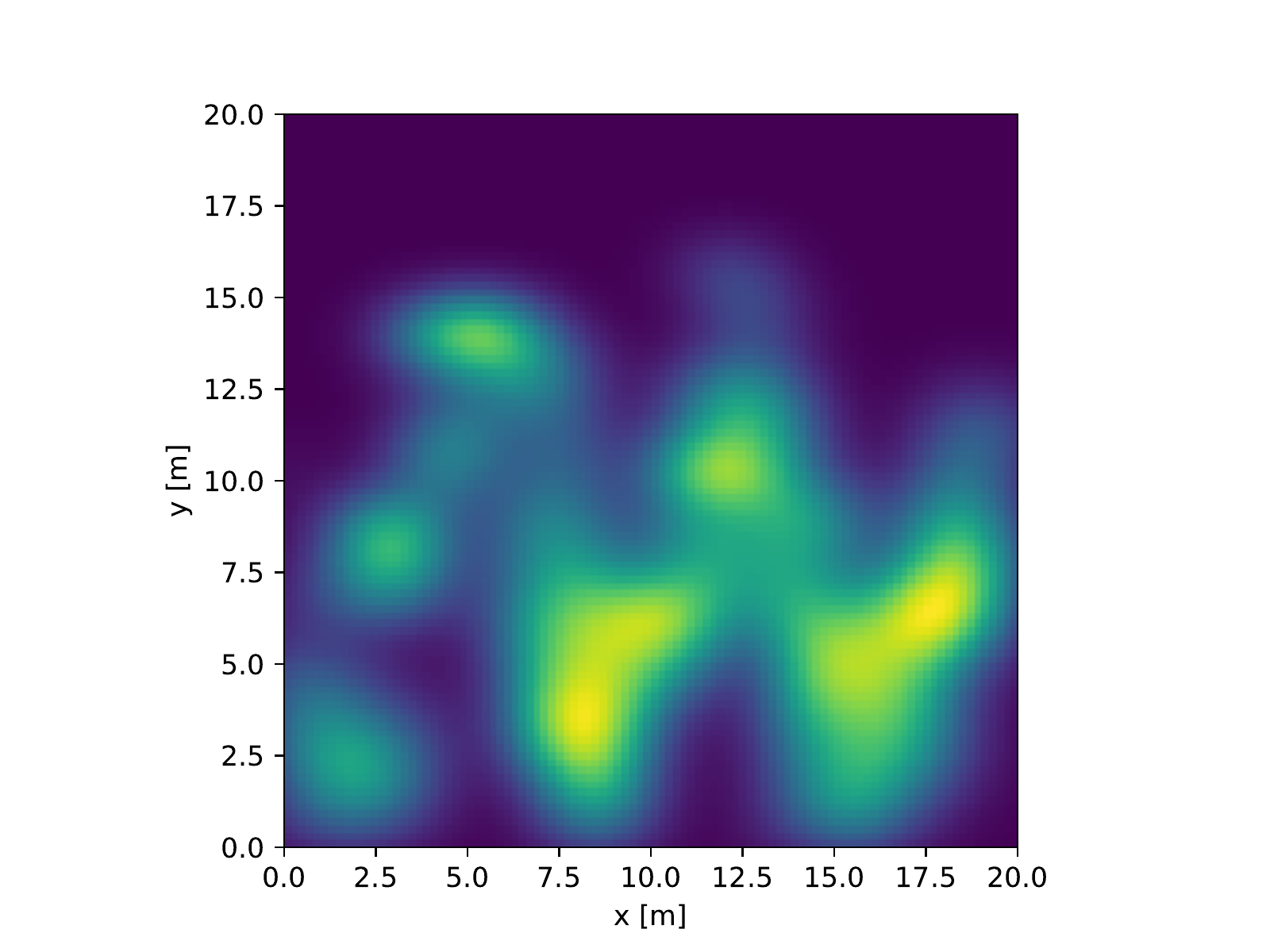} &
				\includegraphics[trim={2cm 0 2cm 0.5cm}, clip, scale=0.65]{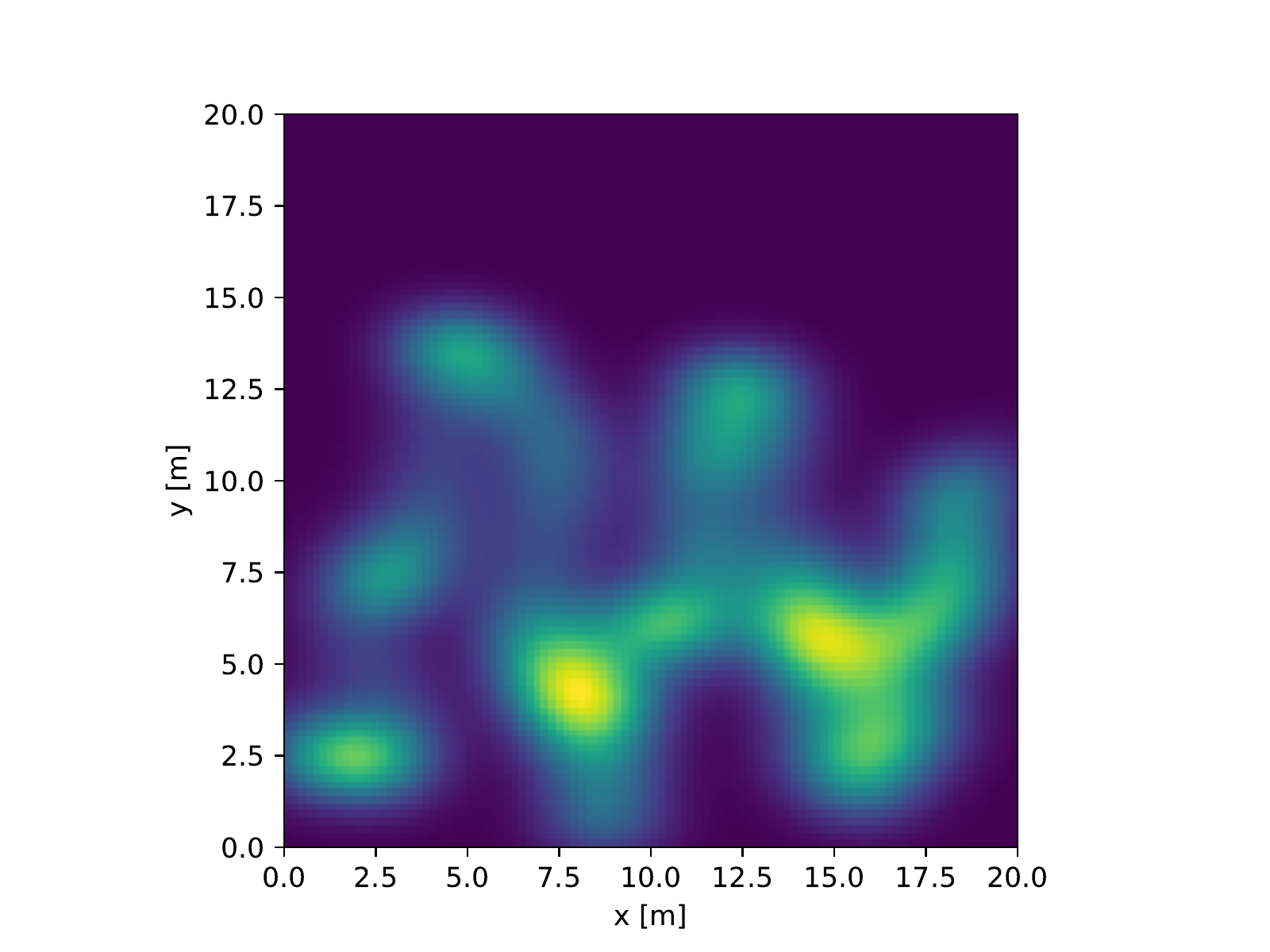} \\
				\rotatebox{90}{\textbf{With additional CTRW physics}}& &
				\includegraphics[trim={2cm 0 2cm 0.5cm}, clip, scale=0.65]{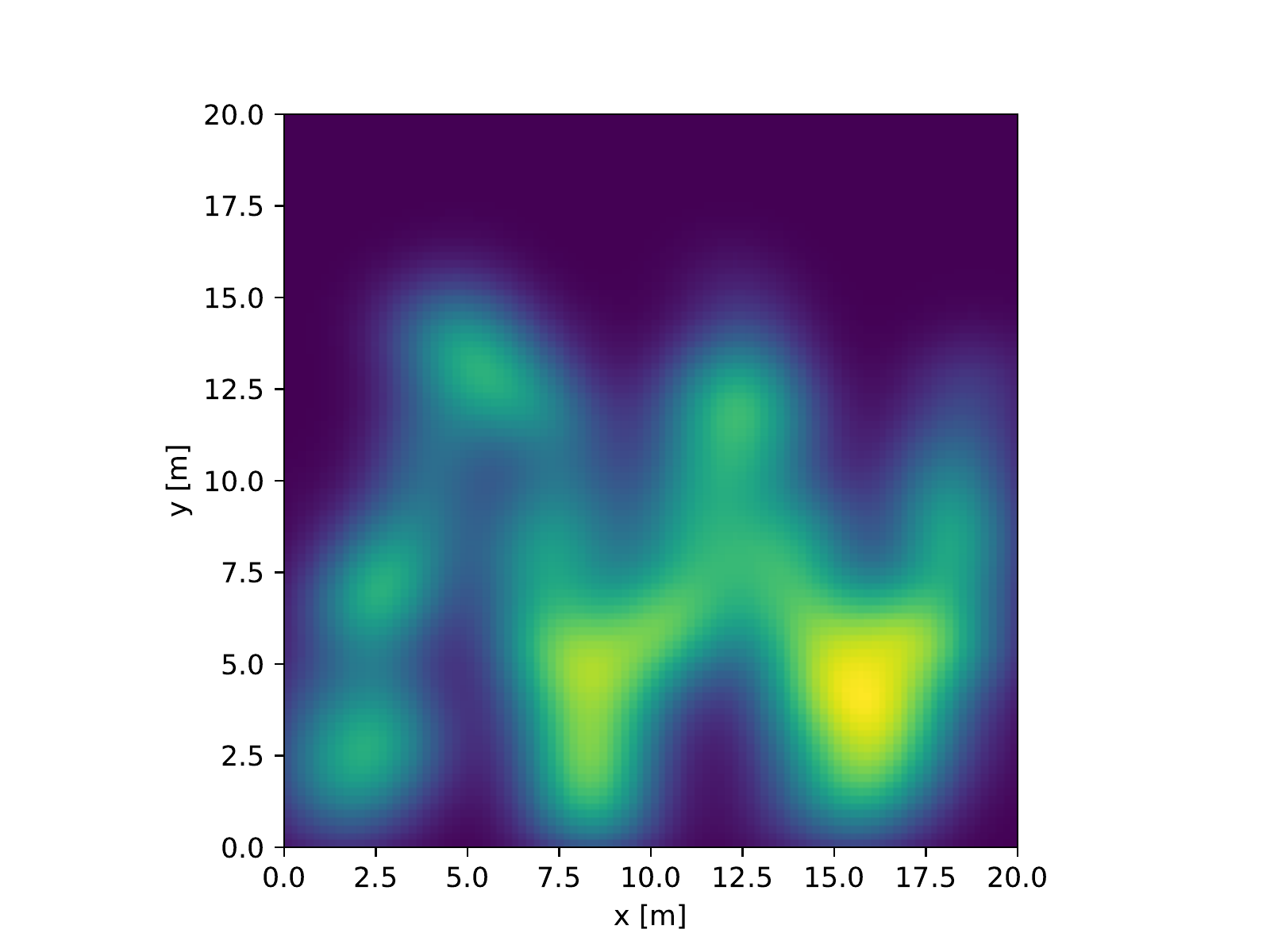}
			\end{tabular}
			\caption{Plumes generated by kernel density estimation from the positions of simulated particles for the three simulation cases; brighter color indicates higher concentration. Plumes in the top row were generated by pure advection plus transverse diffusion, whereas the plume in the bottom row also featured advective scattering from inverse Gaussian $f$. The plume in the left column was simulated on the true velocity field, while those on the right were simulated on the smoothed field. The greater similarity between the true plume (top left) and the plume simulated using the smoothed velocity field plus additional CTRW physics (bottom right) versus the plume simulated using the smoothed velocity field alone (top right) is apparent.}
			\label{fig: plume comparison}
		\end{figure}

\begin{figure}
			\centering
			\includegraphics{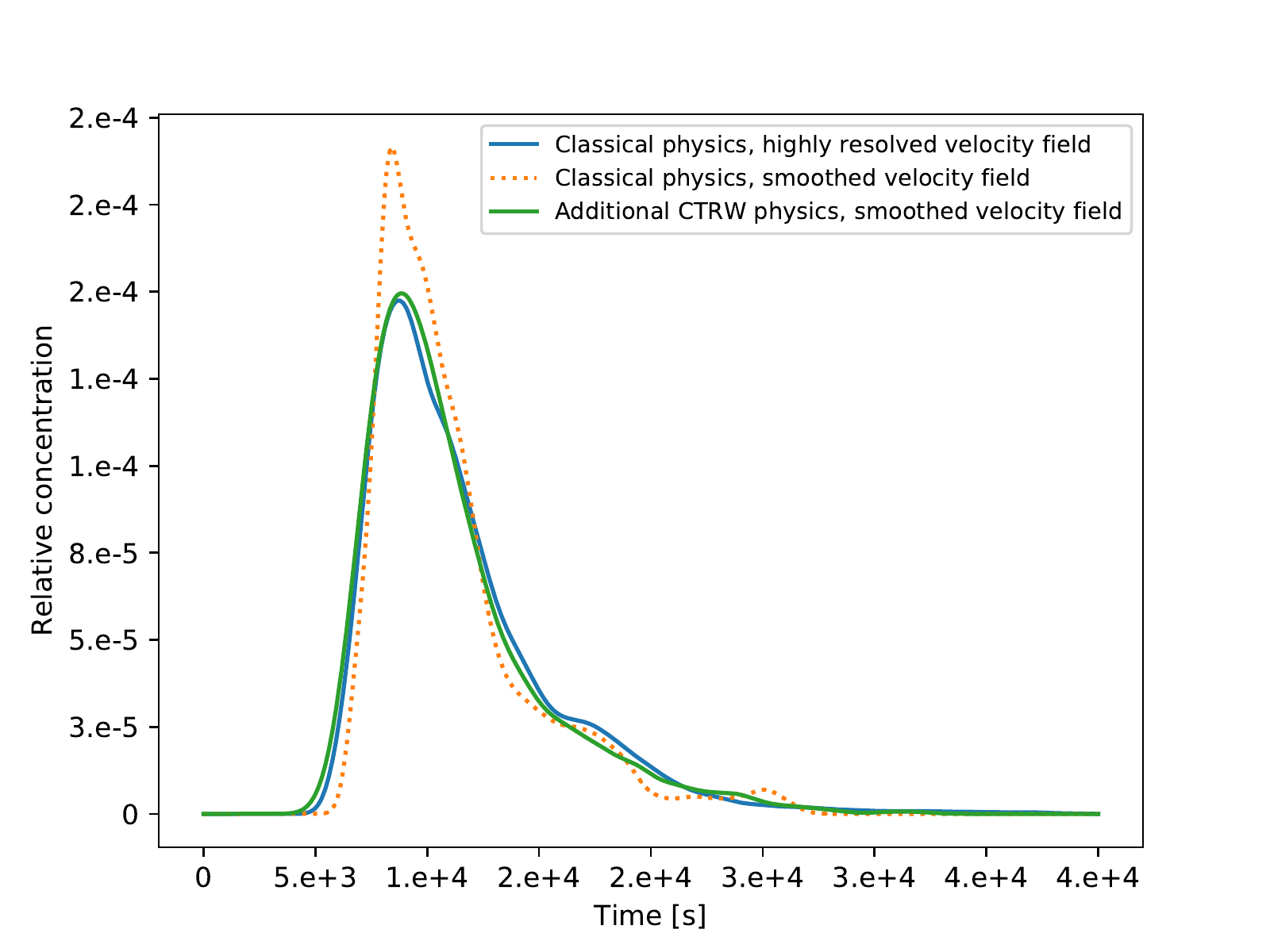}
			\caption{Breakthrough curves at $y=0$ for each of the three simulations.}
			\label{fig: btc comparison}
		\end{figure}

		We see from qualitative comparison of the plume snapshots in Figure \ref{fig: plume comparison} and breakthrough curves in Figure \ref{fig: btc comparison} that the additional CTRW physics (in the form of $f(t)$) do capture much of the transport information that was lost when moving from the highly resolved to the smoothed field. This qualitative assessment is backed by comparison of the 2-norms (also known as Euclidian or Frobenius norms) of the difference between each of the two discretized concentration fields computed using the smoothed velocity field versus the true concentration field. The 2-norm of the error of the concentration field computed without additional CTRW physics is 1.59 times as large as that of the field computed with the additional physics. A similar picture emerges when the $\infty$-norms (largest absolute divergence) are compared. The $\infty$-norm of the no-additional-physics case is 1.68 times as large.  
		
	\section{Summary and conclusion}
		We have presented a hybrid approach to modeling solute transport on velocity fields that are explicitly characterized at a coarse scale, but which contain small-scale physical heterogeneity and/or mass transfer leading to non-Fickian transport. The approach, which we call ``CTRW-on-a-streamline'' is a particle tracking technique that represents an advance of the current state of the art in which, for technical reasons, non-Fickian transport models \textit{or} explicitly-delineated heterogeneous velocity fields are employed, but generally not both. Our approach separates the model parameters characterizing macroscopic flow, subscale advective heterogeneity, and mobile-immobile mass transfer, such that each can be directly specified a priori from available data, where relevant theory exists.
		
		We have presented theory justifying our approach, formalized it in a set of governing equations that illustrate its connection to both classic CTRW and to subordination approaches, and operationalized it: listing an exact algorithm that one can implement. Finally, we presented a numerical demonstration of that implementation that supports the qualitative arguments for capturing small-scale heterogeneous advection validation with the single pdf, $f(r)$.
		
	\section*{Acknowlegments} This paper does not concern a data set. Source code for the simulations is archived at \url{https://doi.org/10.5281/zenodo.3558792}. SKH holds the Helen Unger Career Development Chair in Desert Hydrogeology.
	
	\bibliographystyle{plainnat}
	\bibliography{auroralit, auroralit-rev}

\begin{thebibliography}{45}
\providecommand{\natexlab}[1]{#1}
\providecommand{\url}[1]{\texttt{#1}}
\expandafter\ifx\csname urlstyle\endcsname\relax
  \providecommand{\doi}[1]{doi: #1}\else
  \providecommand{\doi}{doi: \begingroup \urlstyle{rm}\Url}\fi

\bibitem[Banton et~al.(1997)Banton, Delay, and Porel]{Banton1997}
Olivier Banton, Frederick Delay, and Gilles Porel.
\newblock {A New Time Domain Random Walk Method for Solute Transport in 1-D
  Heterogeneous Media}.
\newblock \emph{Groundwater}, 35\penalty0 (6), 1997.

\bibitem[Beaudoin and {De Dreuzy}(2013)]{Beaudoin2013}
A.~Beaudoin and J.~R. {De Dreuzy}.
\newblock {Numerical assessment of 3-D macrodispersion in heterogeneous porous
  media}.
\newblock \emph{Water Resources Research}, 49\penalty0 (5):\penalty0
  2489--2496, 2013.
\newblock ISSN 00431397.
\newblock \doi{10.1002/wrcr.20206}.

\bibitem[Benson and Meerschaert(2009)]{Benson2009}
David~A. Benson and Mark~M. Meerschaert.
\newblock {A simple and efficient random walk solution of multi-rate
  mobile/immobile mass transport equations}.
\newblock \emph{Advances in Water Resources}, 32\penalty0 (4):\penalty0
  532--539, 2009.
\newblock ISSN 03091708.
\newblock \doi{10.1016/j.advwatres.2009.01.002}.

\bibitem[Berkowitz et~al.(2006)Berkowitz, Cortis, Dentz, and
  Scher]{Berkowitz2006}
B.~Berkowitz, A.~Cortis, M.~Dentz, and H.~Scher.
\newblock {Modeling non-Fickian Transport in Geological Formations as a
  Continuous Time Random Walk}.
\newblock \emph{Reviews of Geophysics}, 44:\penalty0 RG2003, 2006.
\newblock ISSN 0004-5411.
\newblock \doi{10.1029/2005RG000178}.

\bibitem[Berkowitz et~al.(2016)Berkowitz, Dror, Hansen, and
  Scher]{Berkowitz2016}
B.~Berkowitz, I.~Dror, S.K. Hansen, and H.~Scher.
\newblock {Measurements and models of reactive transport in geological media}.
\newblock \emph{Reviews of Geophysics}, 54\penalty0 (4), 2016.
\newblock ISSN 19449208.
\newblock \doi{10.1002/2016RG000524}.

\bibitem[Billingsley(1986)]{Billingsley1986}
Patrick Billingsley.
\newblock \emph{{Probability and Measure}}.
\newblock John Wiley {\&} Sons, 2nd edition, 1986.

\bibitem[Bodin(2015)]{Bodin2015}
Jacques Bodin.
\newblock {From analytical solutions of solute transport equations to
  multidimensional time-domain random walk (TDRW) algorithms}.
\newblock \emph{Water Resources Research}, 51\penalty0 (3):\penalty0
  1860--1871, 2015.
\newblock ISSN 19447973.
\newblock \doi{10.1002/2014WR015910}.

\bibitem[Bolster and Dentz(2012)]{Bolster2012}
D.~Bolster and M.~Dentz.
\newblock {Anomalous dispersion in chemically heterogeneous media induced by
  long-range disorder correlation}.
\newblock \emph{Journal of Fluid Mechanics}, 695:\penalty0 366--389, 2012.
\newblock ISSN 00221120.
\newblock \doi{10.1017/jfm.2012.25}.

\bibitem[Carrera et~al.(1998)Carrera, S{\'{a}}nchez-Vila, Benet, Medina,
  Galarza, and Guiner{\`{a}}]{Carrera1998}
Jes{\'{u}}s Carrera, Xavier S{\'{a}}nchez-Vila, Inmaculada Benet, Agust{\'{i}}n
  Medina, Germ{\'{a}}n Galarza, and Jordi Guiner{\`{a}}.
\newblock {On matrix diffusion: Formulations, solution methods and qualitative
  effects}.
\newblock \emph{Hydrogeology Journal}, 6\penalty0 (1):\penalty0 178--190, 1998.
\newblock ISSN 14312174.
\newblock \doi{10.1007/s100400050143}.

\bibitem[Cirpka and Nowak(2003)]{Cirpka2003}
Olaf~A. Cirpka and Wolfgang Nowak.
\newblock {Dispersion on kriged hydraulic conductivity fields}.
\newblock \emph{Water Resources Research}, 39\penalty0 (2):\penalty0 1--12,
  2003.
\newblock ISSN 00431397.
\newblock \doi{10.1029/2001WR000598}.

\bibitem[Comolli et~al.(2016)Comolli, Hidalgo, Moussey, and Dentz]{Comolli2016}
Alessandro Comolli, Juan~J. Hidalgo, Charlie Moussey, and Marco Dentz.
\newblock {Non-Fickian Transport Under Heterogeneous Advection and
  Mobile-Immobile Mass Transfer}.
\newblock \emph{Transport in Porous Media}, 115\penalty0 (2):\penalty0
  265--289, 2016.
\newblock ISSN 15731634.
\newblock \doi{10.1007/s11242-016-0727-6}.

\bibitem[Cortis et~al.(2004)Cortis, Gallo, Scher, and Berkowitz]{Cortis2004}
Andrea Cortis, Claudio Gallo, Harvey Scher, and B.~Berkowitz.
\newblock {Numerical simulation of non-Fickian transport in geological
  formations with multiple-scale heterogeneities}.
\newblock \emph{Water Resources Research}, 40\penalty0 (4):\penalty0 1--16,
  2004.
\newblock ISSN 00431397.
\newblock \doi{10.1029/2003WR002750}.

\bibitem[Cvetkovic and Haggerty(2002)]{Cvetkovic2002}
V.~Cvetkovic and R.~Haggerty.
\newblock {Transport with multiple-rate exchange in disordered media}.
\newblock \emph{Physical Review E}, 65\penalty0 (5):\penalty0 051308, 2002.
\newblock ISSN 1063-651X.
\newblock \doi{10.1103/PhysRevE.65.051308}.

\bibitem[Cvetkovic et~al.(2016)Cvetkovic, Fiori, and Dagan]{Cvetkovic2016}
V.~Cvetkovic, A.~Fiori, and G.~Dagan.
\newblock {Tracer travel and residence time distributions in highly
  heterogeneous aquifers: Coupled effect of flow variability and mass
  transfer}.
\newblock \emph{Journal of Hydrology}, 543:\penalty0 101--108, 2016.
\newblock ISSN 00221694.
\newblock \doi{10.1016/j.jhydrol.2016.04.072}.

\bibitem[Delay and Bodin(2001)]{Delay2001}
Frederick Delay and Jacques Bodin.
\newblock {Time domain random walk method to simulate transport by
  advection-dispersion and matrix diffusion in fracture networks}.
\newblock \emph{Geophysical Research Letters}, 28\penalty0 (21):\penalty0
  4051--4054, 2001.

\bibitem[Dentz and Berkowitz(2003)]{Dentz2003}
Marco Dentz and Brian Berkowitz.
\newblock {Transport behavior of a passive solute in continuous time random
  walks and multirate mass transfer}.
\newblock \emph{Water Resources Research}, 39\penalty0 (5):\penalty0 1--20,
  2003.
\newblock ISSN 00431397.
\newblock \doi{10.1029/2001WR001163}.

\bibitem[Edery et~al.(2014)Edery, Guadagnini, Scher, and Berkowitz]{Edery2014}
Yaniv Edery, Alberto Guadagnini, Harvey Scher, and Brian Berkowitz.
\newblock {Origins of anomalous transport in heterogeneous media: Structural
  and dynamic controls}.
\newblock \emph{Water Resources Research}, 50:\penalty0 1490--1505, 2014.
\newblock ISSN 00431397.
\newblock \doi{10.1002/2013WR015111}.

\bibitem[Fetter(1999)]{Fetter1999}
C.~W. Fetter.
\newblock \emph{{Contaminant Hydrogeology}}.
\newblock Prentice-Hall, 1999.

\bibitem[Gerke and van Genuchten(1993)]{Gerke1993}
H.~H. Gerke and M.~T. van Genuchten.
\newblock {A Dual-Porosity Model for Simulating the Preferential Movement of
  Water and Solutes in Structured Porous Media}.
\newblock \emph{Water Resources Research}, 29\penalty0 (2):\penalty0 305--319,
  1993.

\bibitem[Gotovac et~al.(2009)Gotovac, Cvetkovic, and Andricevic]{Gotovac2009}
Hrvoje Gotovac, Vladimir Cvetkovic, and Roko Andricevic.
\newblock {Flow and travel time statistics in highly heterogeneous porous
  media}.
\newblock \emph{Water Resources Research}, 45\penalty0 (7):\penalty0 W07402,
  2009.
\newblock ISSN 0043-1397.
\newblock \doi{10.1029/2008WR007168}.

\bibitem[Haggerty and Gorelick(1995)]{Haggerty1995}
Roy Haggerty and Steven~M Gorelick.
\newblock {Multiple-rate mass transfer for modeling diffusion and surface
  reactions in media with pore-scale heterogeneity}.
\newblock \emph{Water Resources Research}, 31\penalty0 (10):\penalty0
  2383--2400, 1995.

\bibitem[Haggerty et~al.(2000)Haggerty, McKenna, and Meigs]{Haggerty2000}
Roy Haggerty, Sean~A. McKenna, and Lucy~C. Meigs.
\newblock {On the late-time behavior of tracer test breakthrough curves}.
\newblock \emph{Water Resources Research}, 36\penalty0 (12):\penalty0
  3467--3479, 2000.
\newblock ISSN 00431397.
\newblock \doi{10.1029/2000WR900214}.

\bibitem[Hansen and Berkowitz(2014)]{Hansen2014}
S.~K. Hansen and B.~Berkowitz.
\newblock {Interpretation and nonuniqueness of CTRW transition distributions:
  Insights from an alternative solute transport formulation}.
\newblock \emph{Advances in Water Resources}, 74:\penalty0 54--63, 2014.
\newblock ISSN 03091708.
\newblock \doi{10.1016/j.advwatres.2014.07.011}.

\bibitem[Hansen and Vesselinov(2018)]{Hansen2018a}
S.~K. Hansen and V.~V. Vesselinov.
\newblock {Local Equilibrium and Retardation Revisited}.
\newblock \emph{Groundwater}, 56\penalty0 (1):\penalty0 109--117, 2018.
\newblock ISSN 17456584.
\newblock \doi{10.1111/gwat.12566}.

\bibitem[Hansen et~al.(2016)Hansen, Berkowitz, Vesselinov, O'Malley, and
  Karra]{Hansen2016}
S.~K. Hansen, B.~Berkowitz, V.~V. Vesselinov, D.~O'Malley, and S.~Karra.
\newblock {Push-pull tracer tests: Their information content and use for
  characterizing non-Fickian, mobile-immobile behavior}.
\newblock \emph{Water Resources Research}, 52\penalty0 (12):\penalty0
  9565--9585, 2016.
\newblock ISSN 1093-474X.
\newblock \doi{10.1111/j.1752-1688.1969.tb04897.x}.

\bibitem[Hansen et~al.(2018)Hansen, Haslauer, Cirpka, and
  Vesselinov]{Hansen2018}
S.~K. Hansen, C.~P. Haslauer, O.~A. Cirpka, and V.~V. Vesselinov.
\newblock {Direct Breakthrough Curve Prediction From Statistics of
  Heterogeneous Conductivity Fields}.
\newblock \emph{Water Resources Research}, 54\penalty0 (1):\penalty0 271-- 285,
  2018.
\newblock ISSN 19447973.
\newblock \doi{10.1002/2017WR020450}.

\bibitem[Kang et~al.(2014)Kang, {Le Borgne}, Dentz, Bour, and Juanes]{Kang2014}
P.~K. Kang, T.~{Le Borgne}, M.~Dentz, O.~Bour, and R.~Juanes.
\newblock {Impact of velocity correlation and distribution on transport in
  fractured media: Field evidence and theoretical model}.
\newblock \emph{Water Resources Research}, 51:\penalty0 940--959, 2014.
\newblock \doi{10.1002/2013WR014956.Received}.

\bibitem[Kreft and Zuber(1978)]{Kreft1978}
A.~Kreft and A.~Zuber.
\newblock {On the physical meaning of the dispersion equation and its solutions
  for different initial and boundary conditions}.
\newblock \emph{Chemical Engineering Science}, 33\penalty0 (11):\penalty0
  1471--1480, 1978.
\newblock \doi{10.1016/0009-2509(78)85196-3}.

\bibitem[Margolin et~al.(2003)Margolin, Dentz, and Berkowitz]{Margolin2003}
Gennady Margolin, Marco Dentz, and Brian Berkowitz.
\newblock {Continuous time random walk and multirate mass transfer modeling of
  sorption}.
\newblock \emph{Chemical Physics}, 295:\penalty0 71--80, 2003.
\newblock ISSN 03010104.
\newblock \doi{10.1016/j.chemphys.2003.08.007}.

\bibitem[Michalak and Kitanidis(2000)]{Michalak2000}
Anna~M. Michalak and Peter~K. Kitanidis.
\newblock {Macroscopic behavior and random-walk particle tracking of
  kinetically sorbing solutes}.
\newblock \emph{Water Resources Research}, 36\penalty0 (8):\penalty0
  2133--2146, 2000.
\newblock ISSN 00431397.
\newblock \doi{10.1029/2000WR900109}.

\bibitem[Moslehi and de~Barros(2017)]{Moslehi2017}
Mahsa Moslehi and Felipe P.~J. de~Barros.
\newblock {Uncertainty quantification of environmental performance metrics in
  heterogeneous aquifers with long-range correlations}.
\newblock \emph{Journal of Contaminant Hydrology}, 196:\penalty0 21--29, 2017.
\newblock ISSN 18736009.
\newblock \doi{10.1016/j.jconhyd.2016.12.002}.

\bibitem[Muller and Schuler(2019)]{Muller2019}
Sebastian Muller and Lennart Schuler.
\newblock {GeoStat-Framework/GSTools: Bouncy Blue (v1.0.1)}.
\newblock 2019.
\newblock \doi{10.5281/zenodo.2543658}.

\bibitem[Neretnieks(1980)]{Neretnieks1980}
Ivars Neretnieks.
\newblock {Diffusion in the rock matrix: An important factor in radionuclide
  retardation?}
\newblock \emph{Journal of Geophysical Research: Solid Earth}, 85\penalty0
  (B8):\penalty0 4379--4397, 1980.

\bibitem[Nissan and Berkowitz(2019)]{Nissan2019}
Alon Nissan and Brian Berkowitz.
\newblock {Anomalous transport dependence on P{\'{e}}clet number, porous medium
  heterogeneity, and a temporally varying velocity field}.
\newblock \emph{Physical Review E}, 99\penalty0 (3):\penalty0 1--11, 2019.
\newblock ISSN 24700053.
\newblock \doi{10.1103/PhysRevE.99.033108}.

\bibitem[Oliphant(2007)]{Oliphant2007}
Travis~E. Oliphant.
\newblock {Python for Scientific Computing}.
\newblock \emph{Computing in Science {\&} Engineering}, 9:\penalty0 10--20,
  2007.
\newblock \doi{10.1109/MCSE.2007.58}.

\bibitem[Painter et~al.(2008)Painter, Cvetkovic, Mancillas, and
  Pensado]{Painter2008}
Scott Painter, Vladimir Cvetkovic, James Mancillas, and Osvaldo Pensado.
\newblock {Time domain particle tracking methods for simulating transport with
  retention and first-order transformation}.
\newblock \emph{Water Resources Research}, 44\penalty0 (1):\penalty0 1--11,
  2008.
\newblock ISSN 00431397.
\newblock \doi{10.1029/2007WR005944}.

\bibitem[Reimus and James(2002)]{Reimus2002}
Paul~W. Reimus and Scott~C. James.
\newblock {Determining the random time step in a constant spatial step particle
  tracking algorithm}.
\newblock \emph{Chemical Engineering Science}, 57\penalty0 (21):\penalty0
  4429--4434, 2002.
\newblock ISSN 00092509.
\newblock \doi{10.1016/S0009-2509(02)00396-2}.

\bibitem[Rubin(2003)]{Rubin2003}
Yoram Rubin.
\newblock \emph{{Applied Stochastic Hydrogeology}}.
\newblock Oxford University Press, New York, 2003.

\bibitem[Russian et~al.(2016)Russian, Dentz, and Gouze]{Russian2016}
Anna Russian, Marco Dentz, and Philippe Gouze.
\newblock {Time domain random walks for hydrodynamic transport in heterogeneous
  media}.
\newblock \emph{Water Resources Research}, 52:\penalty0 3309--3323, 2016.
\newblock ISSN 1093-474X.
\newblock \doi{10.1002/ 2015WR018511}.

\bibitem[Schumer et~al.(2003)Schumer, Benson, Meerschaert, and
  Baeumer]{Schumer2003}
Rina Schumer, David~A. Benson, Mark~M. Meerschaert, and Boris Baeumer.
\newblock {Fractal mobile/immobile solute transport}.
\newblock \emph{Water Resources Research}, 39\penalty0 (10):\penalty0 n/a--n/a,
  2003.
\newblock ISSN 00431397.
\newblock \doi{10.1029/2003WR002141}.

\bibitem[Srinivasan et~al.(2010)Srinivasan, Tartakovsky, Dentz, Viswanathan,
  Berkowitz, and Robinson]{Srinivasan2010}
G.~Srinivasan, D.~M. Tartakovsky, M.~Dentz, H.~Viswanathan, B.~Berkowitz, and
  B.~A. Robinson.
\newblock {Random walk particle tracking simulations of non-Fickian transport
  in heterogeneous media}.
\newblock \emph{Journal of Computational Physics}, 229\penalty0 (11):\penalty0
  4304--4314, 2010.
\newblock ISSN 00219991.
\newblock \doi{10.1016/j.jcp.2010.02.014}.

\bibitem[Tsang(1994)]{Tsang1994}
Chin-fu Tsang.
\newblock {Flow channeling in strongly heterogeneous porous media : A numerical
  study}.
\newblock \emph{Water Resources Research}, 30\penalty0 (5):\penalty0
  1421--1430, 1994.

\bibitem[Tyukhova et~al.(2016)Tyukhova, Dentz, Kinzelbach, and
  Willmann]{Tyukhova2016}
Alina Tyukhova, Marco Dentz, Wolfgang Kinzelbach, and Matthias Willmann.
\newblock {Mechanisms of anomalous dispersion in flow through heterogeneous
  porous media}.
\newblock \emph{Physical Review Fluids}, 1\penalty0 (7):\penalty0 1--12, 2016.
\newblock ISSN 2469990X.
\newblock \doi{10.1103/PhysRevFluids.1.074002}.

\bibitem[Uffink et~al.(2012)Uffink, Elfeki, Dekking, Bruining, and
  Kraaikamp]{Uffink2012}
Gerard Uffink, Amro Elfeki, Michel Dekking, Johannes Bruining, and Cor
  Kraaikamp.
\newblock {Understanding the Non-Gaussian Nature of Linear Reactive Solute
  Transport in 1D and 2D: From Particle Dynamics to the Partial Differential
  Equations}.
\newblock \emph{Transport in Porous Media}, 91\penalty0 (2):\penalty0 547--571,
  2012.
\newblock ISSN 01693913.
\newblock \doi{10.1007/s11242-011-9859-x}.

\bibitem[Zhang et~al.(2013)Zhang, Green, and Fogg]{Zhang2013}
Yong Zhang, Christopher~T. Green, and Graham~E. Fogg.
\newblock {The impact of medium architecture of alluvial settings on
  non-Fickian transport}.
\newblock \emph{Advances in Water Resources}, 54:\penalty0 78--99, 2013.
\newblock ISSN 03091708.
\newblock \doi{10.1016/j.advwatres.2013.01.004}.

\end{thebibliography}

\end{document}